\documentclass[iop]{emulateapj}
\usepackage{epsfig}

\usepackage{amsfonts}
\usepackage{amsmath}
\usepackage{mathrsfs}
\usepackage{amssymb}
\usepackage{rotating}

\usepackage{color}
\usepackage{ulem}
\usepackage{xspace}
\usepackage{cancel}
\definecolor{lgray}{gray}{0.85}

\newcommand{\lya} {Ly$\alpha$\xspace}
\newcommand{\civ} {\ion{C}{4}\xspace}
\newcommand{\mgii} {\ion{Mg}{2}\xspace}
\newcommand{\heii}{\ion{He}{2}\xspace}

\newcommand{\unitcgssb}  {erg\,s$^{-1}$\,cm$^{-2}$\,arcsec$^{-2}$\xspace}

\def \cgssb {{\rm\,erg\,s^{-1}\,cm^{-2}\,arcsec^{-2}}}
\def\bea{\begin{eqnarray}}
\def\eea{\end{eqnarray}}

\begin{document}

\title{The Stacked Lyman-Alpha Emission Profile from the Circum-Galactic Medium of $\lowercase{z}\sim2$ 
Quasars\footnotemark[\large $\star$].}\footnotetext[\large $\star$]{
Based on observations obtained at the Gemini Observatory, which is operated by the 
Association of Universities for Research in Astronomy, Inc., under a cooperative agreement 
with the NSF on behalf of the Gemini partnership: the National Science Foundation 
(United States), the National Research Council (Canada), CONICYT (Chile), the Australian 
Research Council (Australia), Minist\'{e}rio da Ci\^{e}ncia, Tecnologia e Inova\c{c}\~{a}o 
(Brazil) and Ministerio de Ciencia, Tecnolog\'{i}a e Innovaci\'{o}n Productiva (Argentina).}
 
\author{Fabrizio Arrigoni Battaia\altaffilmark{1,2}, 
        Joseph F. Hennawi\altaffilmark{2}, 
        Sebastiano Cantalupo\altaffilmark{3},
	J. Xavier Prochaska\altaffilmark{4,5} 
	}
\altaffiltext{1}{European Southern Observatory, Karl-Schwarzschild-Str. 2, D-85748 Garching bei M\"unchen, Germany; farrigon@eso.org}
\altaffiltext{2}{Max-Planck-Institut f\"ur Astronomie, K\"onigstuhl 17, D-69117 Heidelberg, Germany}
\altaffiltext{3}{Institute for Astronomy, Department of Physics, ETH Zurich, CH-8093 Zurich, Switzerland}
\altaffiltext{4}{Department of Astronomy and Astrophysics, University of California, 1156 High Street, Santa Cruz, California 95064, USA}
\altaffiltext{5}{University of California Observatories, Lick Observatory, 1156 High Street, Santa Cruz, California 95064, USA}

\slugcomment{Submitted to ApJ}
\shorttitle{CGM Profile of $\lowercase{z}\sim2$ Quasars.}
\shortauthors{Arrigoni Battaia et al.}
 
\begin{abstract}

In the context of the FLASHLIGHT survey, we obtained deep narrow band images of 15 $z\sim2$ quasars
with GMOS on Gemini-South
in an effort to measure \lya emission from circum- and inter-galactic gas on scales of 
hundreds of kpc from the central quasar.
We do not detect bright giant \lya nebulae (SB $\sim 10^{-17}$ \unitcgssb at distances $>50$~kpc) around any of our sources,
although we routinely ($\simeq 47$\%) detect smaller scale $< 50$~kpc \lya emission at this surface brightness
level emerging from either the extended narrow emission line regions powered by the quasars or by star-formation in their
host galaxies. 
We stack our 15 deep images to study the average extended \lya surface
brightness profile around $z\sim2$ quasars, carefully PSF-subtracting
the unresolved emission component and paying close attention to
sources of systematic error.  Our analysis, which achieves an
unprecedented depth, reveals a surface brightness of SB$_{{\rm
    Ly}\alpha}\sim10^{-19}$ \unitcgssb at $\sim200$~kpc, with a
$2.3\sigma$ detection of \lya emission at SB${_{\rm
    Ly\alpha}=(5.5\pm3.1)\times10^{-20}}$~erg\,s$^{-1}$\,cm$^{-2}$\,arcsec$^{-2}$
within an annulus spanning $50$~kpc~$<R<500$~kpc from the quasars.
Assuming this \lya emission is powered by fluorescence from highly ionized gas
illuminated by the bright central quasar, we deduce an average volume
density of $n_{\rm H}=0.6\times10^{-2}$~cm$^{-3}$ on these large
scales.  Our results are in broad agreement with the densities
suggested by cosmological hydrodynamical simulations of massive
($M\simeq 10^{12.5}$) quasar hosts, however they indicate that the
typical quasars at these redshifts are surrounded by gas that is a 
factor of $\sim 100$ times less dense than the ($\sim 1~{\rm cm^{-3}}$) gas responsible
for the giant bright Ly$\alpha$ nebulae around quasars recently discovered by
our group. 
\end{abstract}

\maketitle

\section{Introduction}

Over the past decade galaxy formation studies have increasingly
emphasized the need to understand the gaseous phases that
surround the galaxies themselves (e.g.,
\citealt{Keres2005,Ocvirk2008,Fumagalli2011,Stinson2012,Fumagalli2014,Faucher-Giguere2015}),
which represent both the fuel for and product of star-formation.
Specifically, the circum-galactic medium (CGM), which represents gas
extending from $\simeq 20-300$ kpc from galaxies, encodes information
about the complex interplay between outflows from galaxies and
accretion onto them from the intergalactic medium (IGM) (e.g.,
\citealt{Fraternali2015, Nelson2015}).  The CGM has been
typically studied by analyzing absorption features along
background sightlines at small impact parameter from galaxies
\citep{Croft2002, Bergeron2004, Rudie2012, Crighton2015, Nielsen2015, Crighton2013} and
quasars
\citep{Bowen2006, Hennawi2006,Hennawi2007,Prochaska2009,Hennawi2013,Farina2013,Farina2014,Prochaska2013,Prochaska2013b,Johnson2015}.

Specifically, using this technique with projected quasar pairs, it has
been shown that massive galaxies hosting quasars at $z\sim2$ are
surrounded by a massive ($M_{\rm CGM}>10^{10}$~M$_\odot$), enriched
($Z\gtrsim0.1$~Z$_{\odot}$), cool ($T\approx10^4$~K) CGM with high
covering factor \citep[$\approx60$~\% within the virial radius of
  160~kpc;][]{Prochaska2013,Prochaska2013b,Prochaska2014}.
Notwithstanding great effort, state-of-the-art simulations of galaxy
formation still struggle in reproducing these observations
(\citealt{Rahmati2013,Fumagalli2014,Faucher-Giguere2015, Meiksin2015},
but see also \citealt{Rahmati2015,Faucher-Giguere2016}).
Nevertheless, these absorption studies are limited by the paucity of bright background sources and the inherently
one-dimensional nature of the technique, and hence do not paint a complete picture of the CGM.
Direct observations of the CGM in emission would be highly complementary, and this new observable would
enable a more constrained comparison to the results of simulations.

In particular, it has been suggested that the \lya transition should
be the easiest line to detect from the quasar CGM, and  
from the IGM on larger scales. Indeed, gas in the
CGM and IGM could reprocess impinging UV radiation
resulting in fluorescent \lya emission.  Although the fluorescence
signal powered by the ambient UV background (\citealt{Hogan1987,
  Binette1993, Gould1996, Cantalupo2005}), with an expected surface
brightness (SB) SB$_{{\rm Ly}\alpha}\sim10^{-20}$ \unitcgssb (e.g.,
\citealt{Rauch2008}), is still too faint to be detectable,
this emission can be boosted up to observable levels by the
additional ionizing flux of a nearby quasar (\citealt{Rees1988, HaimanRees2001, Alam2002,Cantalupo2005}).

Given the extreme challenge of detecting this faint quasar powered
fluorescent emission, previous searches have been difficult to interpret
(e.g., \citealt{Fynbo1999, Francis2004, Cantalupo2007, Rauch2008}).
The most compelling candidates for quasar powered Ly$\alpha$
fluorescence are the compact sources, {\it a.k.a.} ``dark galaxies'',
at Mpc distances from quasars discovered by \citet{Cantalupo2012}.
Indeed, they reported an excess of \lya emitters (LAEs) with
rest-frame \lya equivalent width exceeding the maximum expected from
star-forming regions, i.e. ${\rm EW}_0^{{\rm Ly}\alpha}>240$\AA\ (e.g.,
\citealt{Charlot1993}).

At smaller distances from the quasar $r \lesssim 500\,{\rm kpc}$, in
the CGM and at the CGM-IGM interface, one also expects gas to be
illuminated. Motivated by this, many have searched for Ly$\alpha$
emission close to the quasar (e.g., \citealt{HuCowie1987,
  Heckman1991spec, Heckman1991, Christensen2006, Hennawi2009,
  North2012, Hennawi2013, Roche2014}). While the consensus emerging
from this work is a relatively high $50-70\%$ rate of detection
of extended \lya emission on small-scales (radius of $10-50$ kpc), convincing
evidence for larger scale emission remains elusive.

Recently more systematic searches with higher sensitivity have uncovered several
very extended \lya nebulae around $z\sim2$ QSOs
(\citealt{Cantalupo2014, Jackpot, Trainor2013,Martin2014a,Martin2015}).
In particular, our group discovered the two most striking examples of
giant \lya nebulae around the radio-quiet quasars: UM~287
(\citealt{Cantalupo2014}), and SDSSJ0841+3921 (\citealt{Jackpot}).  With
respective sizes of 460 kpc and 310 kpc, these constitute the largest
nebulae ever observed around a QSO, signifying large reservoirs of
cool ($T\sim10^4$ K) gas.  Their bright \lya emission has been
explained as fluorescent recombination emission from the CGM/IGM
powered by the intense ionizing radiation emitted by the central
quasar. Under this interpretation, independent arguments based on
cosmological zoom-in simulations post-processed with radiative transfer
(\citealt{Cantalupo2014}), as well as the non-detection of \heii and \civ
emission from the nebula (\citealt{FAB2015}), indicate the \lya emission from the
UM~287 nebulae comes from  a large population of
compact ($\lesssim20$~pc), dense ($n_{\rm H}\gtrsim3$~cm$^{-3}$), cool
gas clumps (\citealt{FAB2015}). \citet{Jackpot} came to similar conclusions
about the gas responsible for the \lya emission around SDSSJ0841+3921 based
on joint modeling of the nebula in absorption and emission.

Interestingly, an increasing number of independent observations point
in the same direction, suggesting high densities (comparable to the
interstellar medium) and small cloud sizes in the quasar CGM
(\citealt{Prochaska2009, QPQ8}). However, note that only a fraction of
objects studied through absorption line techniques require high
ISM-like densities (e.g., \citealt{Prochaska2009, QPQ8}), although several
assumptions are required in order to constrain this parameter
(e.g. photoionization models).
The densities found in giant \lya nebulae are orders of magnitude higher than predictions 
from simulations on CGM scales ($n_{\rm
  H}\sim10^{-2}-10^{-3}$~cm$^{-3}$; e.g., \citealt{Rosdahl12}),
probably reflecting the inability of current simulations of galaxy
formation to resolve small scale structure in the cool CGM
(\citealt{Crighton2015, FAB2015, Jackpot}).

High density clumps have also been invoked to explain the extended
emission line regions (EELR; e.g., \citealt{Stockton2006}) around
active galactic nuclei (AGN; \citealt{Stockton2002,
  FuStockton2007}). The narrow lines ($\lesssim
1000$~km~s$^{-1}$) around AGN can extend out to tens of kpc, and are
thought to be mainly powered by the AGNs (e.g.,
\citealt{Husemann2014}).  Given their high surface brightness
$\sim10^{-16}$~\unitcgssb, the 
searches for \lya emission around quasars clearly detect the \lya signal from these regions (e.g.,
\citealt{Christensen2006}).

Although the discovery of bright high surface brightness \lya nebulae on large scales
around quasars provides an important new diagnostic of the CGM, revealing
problems with our understanding of cool gas in massive halos, 
they represent a relatively rare phenomenon and it is unclear whether the
extreme properties implied by these nebulae are representative of the quasar CGM
as a whole.
Indeed, as summarized in \citet{Jackpot}, 
three independent surveys (\citealt{FAB2013, Hennawi2013, Trainor2013}) suggest that only $\approx$10~\% of 
quasars exhibit giant \lya nebulae, i.e. with high SB emission (${\rm SB}_{\rm Ly\alpha}\sim10^{-17}$~\unitcgssb
at distances $>50$~kpc from the quasar). It is still unclear whether this frequency of detection results from different physical
properties of the surrounding gas (e.g. higher densities in the giant \lya nebulae), 
and/or variations in the solid angle into which quasars emit their ionizing radiation
(\citealt{Cantalupo2014, Jackpot}), and/or peculiar environment of these sources (\citealt{Jackpot}).

In this paper we conduct a stacking analysis of
narrow-band (NB) images targeting the \lya line for 15 $z\sim 2$ QSOs that do
{\it not} show giant \lya nebulae. Our goal is to measure the average surface
brightness  of the quasar CGM, an important quantity for planning future
integral field unit (IFU)
observations with new instruments such as the Multi Unit Spectroscopic Explorer (MUSE; \citealt{Bacon2010}) 
and the Keck Cosmic Web Imager (KCWI; \citealt{Morrissey2012}). 
These data have been collected in the framework of the larger
Fluorescent Lyman-Alpha Survey of cosmic Hydrogen iLlumInated by
hIGH-redshifT quasars (FLASHLIGHT, see also \citealt{FAB2013}), which
aims to better constrain the fluorescence signal on both CGM and IGM
scales around $z\sim2-3$ QSOs.  This survey has been conducted using
custom-built narrow-band filters with a program using the Gemini
Multi-Object Spectrograph (GMOS, \citealt{Hook2004}) on Gemini-South,
and a second one using the Low-Resolution Imaging Spectrometer (LRIS,
\citealt{Oke1995}) on Keck.  While the on-going program on Keck
resulted so far in the discovery and study of the aforementioned giant
\lya nebulae around UM~287 (\citealt{Cantalupo2014}) and
SDSSJ0841+3921 (\citealt{Jackpot})\footnote{Note that this nebula was
  previously detected with long-slit observations by
  \citet{Hennawi2013}.}, in this work we focus on the data collected
at Gemini-South.

To our knowledge, similar \lya stacking analysis has never been
conducted around QSOs, but have been explored in the case of LAEs and
Lyman break galaxies (LBGs), which exhibit extended \lya halos on
scales of tens of kpc.  In particular, \citet{Rauch2008} showed that
the average and median SB$_{{\rm Ly}\alpha}$ profiles for 27 line
emitters at $2.66<z<3.75$ extend out to $\sim50$~kpc at a level of
SB$_{{\rm Ly}\alpha}\sim10^{-19}$ \unitcgssb.  After this initial
study of individual objects based on a 92 hour ultra-deep longslit
observation, subsequent workers chose to instead stack the data of
several objects observed with much shorter exposure
times. Specifically, \citet{Steidel2011} reported evidence for
extended \lya emission out to $\sim80$~kpc at a similar SB level
around $z=\langle 2.65 \rangle$ LBGs by stacking 92
objects. \citet{Matsuda2012} performed a stacking analysis for
$\sim2000$ LAEs and 24 LBGs, confirming the result of
\citet{Steidel2011} in the case of LBGs, and showing a probable
environmental dependence of the size and luminosities of their average
LAE \lya profiles, with larger and brighter \lya halos in the richest
environment, though always fainter than LBGs.  
\citet{Momose2014} expanded the work
by \citet{Matsuda2012} up to $z\sim6$ also detecting extended \lya
halos.  Recently \citet{Wisotzki2016}, exploiting the higher
sensitivity of the MUSE instrument
(\citealt{Bacon2010}), showed that the \lya SB profiles of
individual high-$z$ galaxies can be studied down to SB$_{{\rm
    Ly}\alpha}\sim10^{-19}$ \unitcgssb (on scales of tens of kpc) if
very long 27 hour exposure times are employed.

Note that several mechanisms can in principle power extended \lya
halos, making it difficult to provide a unique interpretation,
especially when only \lya images are available.  The main mechanisms
invoked in the literature, which may even act together include: (i)
resonant scattering of \lya line photons
(\citealt{Dijkstra2008,Steidel2011,Hayes2011}), (ii) photoionization
by stars or by a central AGN (\citealt{Geach2009,Overzier2013}), (iii)
cooling radiation from cold-mode accretion (e.g.,
\citealt{Fardal2001,Yang2006,Faucher2010,Rosdahl12}), and (iv)
shock-heated gas by galactic superwinds (\citealt{Taniguchi2001}).  An
exemplar case for the ambiguity that can arise in interpreting \lya
emission on $\sim 100~{\rm kpc}$ scales is the ongoing debate about the
nature of the \lya blobs (LABs), i.e. large ($50-100$~kpc)
luminous ($L_{\rm Ly\alpha}\sim10^{43-44}$~erg~s$^{-1}$) \lya nebulae
at $z\sim2-6$ (e.g., \citealt{Steidel2000,
  Matsuda2004,Dey2005,Prescott2009,Yang2011, FAB2015LAB}) -- although
there is increasing evidence that LABs are frequently associated with
obscured AGN (\citealt{Geach2009,Overzier2013,Prescott2015}). While the
brightest giant Ly$\alpha$ nebulae discovered around luminous quasars
appear to be unambiguously powered by photoionization from the central source \citep{Cantalupo2014, Jackpot},
at the much lower SB$_{\rm Ly\alpha}\sim 10^{-19}-10^{-20}$ \unitcgssb levels that we probe in this study, other
emission mechanisms are also plausible. 
In this work we focus mainly on the photoionization scenario, however additional data and a detailed comparison
with cosmological simulations of very massive systems are needed to assess the contribution from the different powering mechanisms.

This paper is organized as follows.  We summarize our observations in
\S\ref{sec:data}.  In \S\ref{sec:datared}, we describe the data reduction
procedures, the global surface brightness limits of our data, the
masking technique used before the SB profile estimate, and the
characterization of the point-spread-function (PSF).  In
\S\ref{sec:results} we present the stacking analysis around the 15
QSOs in our sample and the resulting extended \lya SB profile.
In \S\ref{sec:Monte Carlo} we test the
significance of our results, while in \S\ref{sec:lastDisc} we interpret 
our results in the context of the simple model for emission from the CGM
presented in \citet{Hennawi2013}, and compare to other measurements in the
literature.  Finally, we summarize and conclude in \S\ref{sec:Conclusion}.

Throughout this paper, we adopt the cosmological parameters $H_0 = 70$
km s$^{-1}$ Mpc$^{-1}$, $\Omega_M=0.3$ and $\Omega_{\Lambda}=0.7$.  In
this cosmology, 1\arcsec\ corresponds to 8.2 physical kpc at
$z=2.253$. All magnitudes are in the AB system (\citealt{Oke1974}),
and all distances are proper.

\section{Observations}
\label{sec:data}

In March 2013 we successfully installed a custom
narrow-band filter for the Gemini Multi
Object Spectrograph (GMOS, \citealt{Hook2004}) on the Gemini South
telescope, targeting \lya emission around quasars at ${z=2.253}$
(${\lambda_{\rm center}=3955}$~\AA, ${{\rm FWHM}=32.7}$~\AA, peak
transmission T$_{\rm peak}=53.88$\%)\footnote{The filter is still
  available at the telescope for observations.  Note that in June
  2014, after our last run, the CCDs of GMOS-S were upgraded to the
  new Hamamatsu CCDs, more sensitive at longer wavelengths,
  e.g. $\lambda>5000$~\AA. For this reason, the overall efficiency of
  the system at 4000~\AA\ has been degraded by 25\% (for the blue
  sensitive CCD).  }.
In one year we observed a total of 17 quasars.  Three of these have
longer integrations, typically of 5 hours in a series of
dithered\footnote{For all our observations executed a random dithering
  pattern within a 15\arcsec$\times$15\arcsec\ box, in order to fill
  the two CCD gaps of $\sim2.8$ arcsec each.} 1800s exposures,
achieving a depth of SB$_{{\rm Ly}\alpha}\sim2\times10^{-18}$
\unitcgssb ($1\sigma$ in 1 square arcsec). The other 14 quasars were
observed in a fast survey mode, to try to uncover new giant \lya
nebulae similar to UM~287 (\citealt{Cantalupo2014}) and SDSSJ0841+3921
(\citealt{Jackpot}).  The `fast survey' observations were carried out
using typical exposure times of $\sim2$ hours in a series of dithered
1200s exposures, achieving an average depth of SB$_{{\rm
    Ly}\alpha}\sim4.5\times10^{-18}$ \unitcgssb ($1\sigma$ in 1 square
arcsec).

To study a representative sample of quasars we selected our targets
from the SDSS/BOSS catalogue (\citealt{Paris2014}). In particular, we
focus on the brightest quasars within our custom narrow-band filter
which do not have bright stars in their vicinity, or significant
Galactic extinction in their direction.  As NIR spectra are not
available for most of the SDSS QSOs, accurate redshifts are determined using 
the \mgii line, which has a small and known offset from the systemic redshift of the quasar (\citealt{Richards2002, Shen2016}).
Indeed, the typical
uncertainty on these redshift estimates is $\sigma_{z}\sim0.003$ (or
equivalently $\sim270$~km~s$^{-1}$), which is much smaller than the
width of the narrow-band filter used, i.e. $\Delta z=0.027$ (or
equivalently $\Delta v=2479$~km~s$^{-1}$). Thus, to be sure that the
\lya emission of our targets falls within the narrow band filter, we
selected only QSOs whose redshift gives a maximum shift of
$\pm5$~\AA\ from the filter's center (or equivalently $\delta v =
370$~km~s$^{-1}$).  In the Appendix we quantify the flux losses due to
the uncertainty on the systemic redshift of our sample to be of the
order of 3\%, and thus negligible.

In addition to the NB data, we have observed the same fields in
$g$-band.  For the long integrations the typical exposure time was 3
hours in a series of dithered 240s exposures, whereas in the fast
survey we observed for 40 minutes in a series of dithered 300s
exposures.  The `fast-survey' data were collected in March 2014
in a 4 night run (program ID: {\tt GS-2014A-C-2}), while the longer
exposures were obtained in service mode during 2013-2014 (program ID:
{\tt GS-2013A-Q-36}).The seeing was $\sim$0.7\arcsec\ for the service mode program, while
it ranged between 0.5\arcsec\ and 1.9\arcsec\ during the visitor run,
with a median seeing of $\sim$~1.2\arcsec.  A 4 night run (program ID:
{\tt GS-2013B-C-4}) was completely lost in September 2013, due to bad
weather conditions.
The observations taken with GMOS/Gemini-South and used in
this work are summarized in Table \ref{Tab:GMOS}.

\begin{deluxetable*}{lccccc}
\tablewidth{0pt}
\tabletypesize{\small}
\tablecaption{Summary of the GMOS/Gemini-South Observations used in this work.}
\tablehead{
\colhead{Target                 }&
\colhead{$z$               }&
\colhead{$i$-mag      }&
\colhead{Exp. Time NB$^{\rm a}$       }&
\colhead{Exp. Time $g^{\rm b}$}&
\colhead{Depth$^{\rm c}$                    }\\
\colhead{                      }&
\colhead{                      }&
\colhead{(AB)                      }&
\colhead{(hours)                 }&
\colhead{(hours)                 }&
\colhead{(cgs/arcsec$^2$)              }\\
}
\startdata
SDSSJ081846.64+043935.2&  2.255	&   19.38        &      2        & 	  0.5	    &	$3.8\times10^{-18}$   \\
SDSSJ082109.79+022128.4&  2.254	&   18.91	 &	2	 &	  0.5	    &	$3.8\times10^{-18}$   \\
SDSSJ084117.87+093245.3&  2.254	&   18.02	 &	5	 &	  3.0	    &	$2.5\times10^{-18}$   \\
SDSSJ085233.00+082236.2&  2.253	&   18.31	 &	2	 &	  0.7	    &	$4.4\times10^{-18}$   \\
SDSSJ093849.67+090509.7&  2.255	&   17.33	 &	5	 &	  3.0	    &	$2.1\times10^{-18}$   \\
SDSSJ100412.88+001257.6&  2.253	&   18.62	 &	2	 &	  0.5	    &   $5.0\times10^{-18}$   \\
SDSSJ104330.46-023012.6&  2.252 &   18.37	 &	2        &	  0.5	    &	$4.8\times10^{-18}$   \\
SDSSJ110650.53+061049.9&  2.253 &   18.74    	 &	1.7	 &	  0.4	    &	$6.1\times10^{-18}$   \\
SDSSJ113240.86-014818.9&  2.251 &   19.19    	 &	2	 &	  0.7	    &	$5.2\times10^{-18}$   \\
SDSSJ121503.13+003450.6&  2.255 &   19.36    	 &	2	 &	  0.7	    &	$4.9\times10^{-18}$   \\
SDSSJ131433.84+032322.0&  2.251 &   18.52  	 &	1.7	 &	  0.4	    &   $5.8\times10^{-18}$   \\
SDSSJ141027.12+024555.8&  2.252	&   19.31	 &	2	 &	  0.5	    &	$4.9\times10^{-18}$   \\
SDSSJ141936.61+045430.8&  2.254	&   19.51	 &	1.7	 &	  0.4	    &	$4.6\times10^{-18}$   \\
SDSSJ151521.88+070509.8&  2.254	&   19.96	 &	2	 &	  0.3       &	$4.2\times10^{-18}$   \\
SDSSJ160121.02+064530.3&  2.257	&   19.45	 &	1.7	 &	  0.5	    &	$4.5\times10^{-18}$   \\
\enddata
\tablenotetext{a}{Total exposure time for the observations with the narrow-band filter.} 
\tablenotetext{b}{Total exposure time for the observations in the $g$-band.} 
\tablenotetext{c}{1$\sigma$ surface brightness limit [\unitcgssb] in 1 square arcsec for the NB images (see \S\ref{sec:datared}).}
\label{Tab:GMOS}
\end{deluxetable*}

\section{Data Reduction}
\label{sec:datared}

As our goal is to detect extremely low surface brightness emission for which control
of systematics is crucial,  we opted
to write our own custom data-reduction pipeline in the Interactive Data Language (IDL),
and used only the {\tt gmosaic} routine in the publicly available Gemini/IRAF GMOS package
\footnote{http://www.gemini.edu/sciops/data-and-results/processing-software},
to correctly mosaic the three different chips of the CCD. After mosaicing, we used
our custom pipeline to bias-subtract the images, and flat-field them using
twilight flats. To improve the flat-fielding, essential for detecting
faint extended emission across the fields, we further corrected for the
illumination patterns using night-sky flats.
The night-sky flats were produced by combining the unregistered
science frames with an average sigma-clipping algorithm after masking
out all the objects. Satellite trails, CCD edges, bad pixels, and
saturated pixels were masked. In particular, we created a bad-pixel
mask using an average dark frame obtained from dark images with the
same exposure time and binning as our science frames. We found that
creating a custom bad-pixel mask was necessary because of the
increased number of bad pixels arising from our non-standard binning
of $4\times 4$ pixels (resulting in a pixel scale of
0.29\arcsec~pixel$^{-1}$). Given the low count levels through our NB
filter, this binning was necessary to minimize read-noise, resulting
in nearly background limited observations for our choice of NB
exposure time ($\geq1200$~s, see \S\ref{sec:data}).  Each individual
frame was cleaned from cosmic rays using the L.A.Cosmic algorithm
(\citealt{vanDokkum2001}).

The final stacked images in each filter (NB and $g$-band) were
obtained by averaging the science frames with ${\rm S\slash N}$
weights and masking pixels rejected via a sigma clipping algorithm.
Sky-subtraction was performed on the individual frames before the
final combine. To ensure that we do not mistakenly subtract any
extended emission, or introduce systematics during the sky
subtraction, we simply estimated an average sky value for each
individual image, after masking out all sources, and subtracted this
constant value from each image.  Given the very small field
distortions of GMOS\footnote{The field distortions of GMOS-S are
  regularly checked. For the E2V detector they were estimated to be 1
  pixel in the x-direction and 2 pixels in the y-direction at the edge
  of the field at the time of observations (Pascale Hibon, private
  communication). Note that these values correspond to the unbinned
  CCD, and thus 1 pixel corresponds to 0.07\arcsec, resulting
  in distortions significantly smaller than the pixel scale
  of our $4\times 4$ binned images (0.29\arcsec~pixel$^{-1}$). We
  confirmed this small level of distortion by comparing the SDSS-DR9
  catalogue with stars in our individual frames.}, we calibrated the
astrometry after the stacking\footnote{In this way, we avoided the
  challenge of finding a good astrometric solution for individual
  shallow NB frames with a limited number of sources.}.  The
astrometry was calibrated with the SDSS-DR9 catalogue using {\tt
  SExtractor} (\citealt{Bertin1996}) and {\tt SCAMP}
(\citealt{Bertin2006}). The RMS uncertainties in our astrometric
calibration are $\sim$0.2\arcsec.

For the flux calibration of the NB imaging, we use the standard stars
(H600, G60-54, G138-31) that were repeatedly observed during the
observations, typically at the beginning and at the end of the
night. To avoid systematics in the flux calibration, these
spectrophotometric stars are selected to be free of any feature at the
wavelength of interest, and to have a good sampling of their tabulated
spectrum (at least 1~\AA).  All data show consistent zero points
during different nights, and are stable during each single night (ZP$_{\rm
  NB}=22.11$).  Regarding the $g$-band images, the flux calibration is
performed using several photometric stars in different PG-fields
(PG0918+029, PG1047, PG1633+099) observed at the beginning and at the
end of each night.  Also in this case, all data show consistent zero
points, with good agreement between different nights, and stability
during each single night (ZP$_{g}=28.45$).  Note that our $g$-band zero point value
is consistent with the GMOS tabulated values\footnote{http://www.gemini.edu/sciops/instruments/performance-monitoring/data-products/gmos-n-and-s/photometric-zero-points}.
The uncertainty in the derived zero-points is $\approx0.03$ mag.

During the data reduction steps (e.g. bad pixel masking, satellite
trails masking, etc.), our pipeline consistently propagates the
errors, and produces a variance image $\sigma^2$ for each stacked
image, which contains contributions from the sky and background photon
counting as well as read noise.
An accurate final noise model
is of fundamental importance for the
measurement of very low surface brightness signals
(\citealt{Hennawi2013,FAB2015LAB,FAB2015}).  We compute a global
surface brightness limit for detecting the \lya line using a global
root-mean-square (rms) of the images.  To calculate the global rms per
pixel, we first mask out all the sources in the images, paying
particular attention to the scattered light and halos of bright
foreground stars, and then compute the standard deviation of sky
regions using a sigma-clipping algorithm. We convert these rms values
into the surface brightness limits {\it per} 1 sq.\,arcsec aperture
listed in Table~\ref{Tab:GMOS}. 

\subsection{Image Preparation for the Profile Extraction}
\label{imPrep}

\begin{figure*}
\centering
\epsfig{file=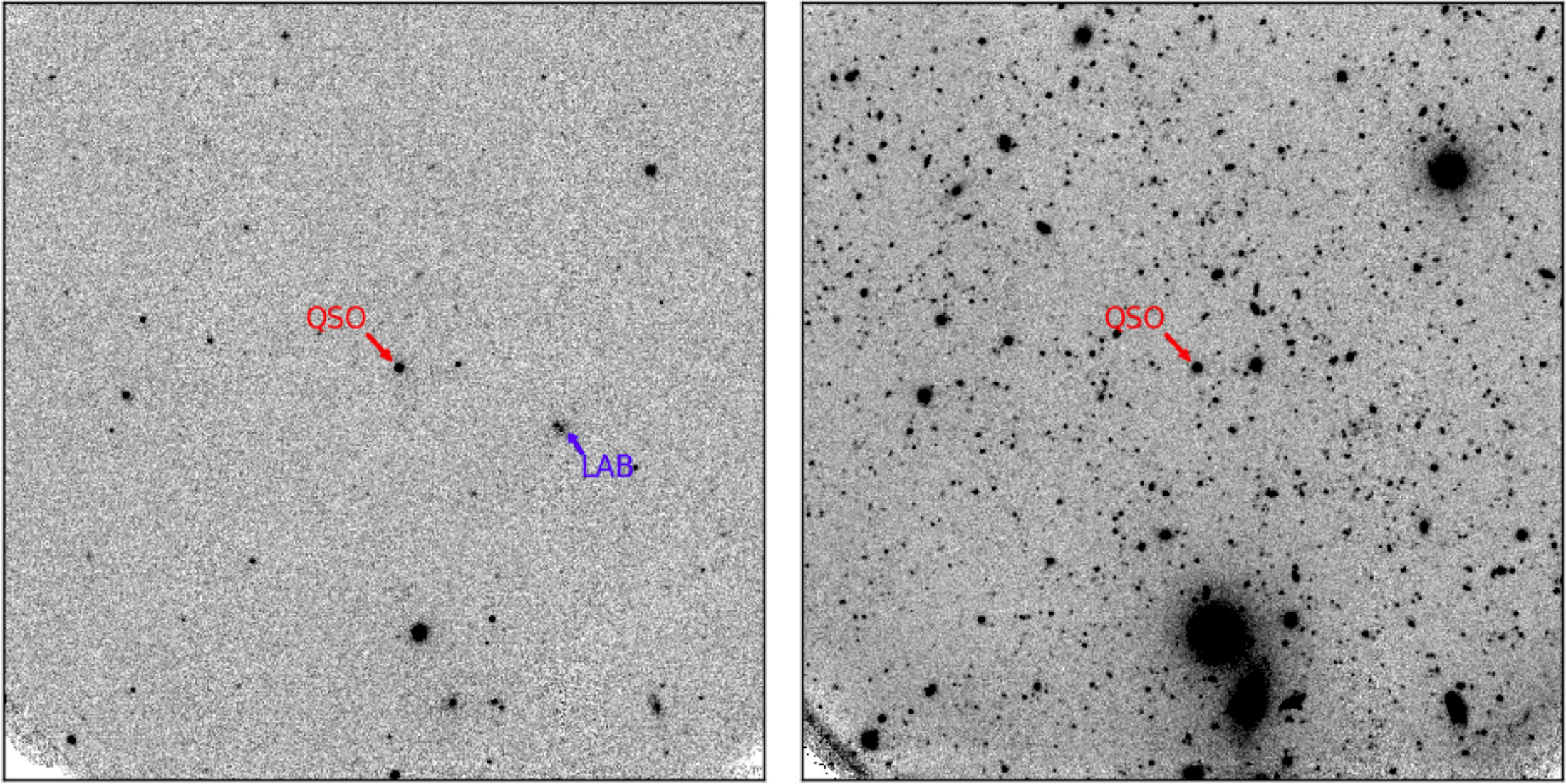, width=0.49\textwidth, angle=270, clip} 
\caption[Example of GMOS/Gemini-South data (FLASHLIGHT survey).]{NB (left) and $g$-band (right) images for the quasar SDSSJ121503.13+003450.6. The images show the complete field of view of the GMOS-S/Gemini instrument, 
i.e. 5.5\arcmin~$\times$~5.5\arcmin. 
The position of the QSO and of the LAB discovered in this field are indicated. Note the flatness of the images, indicating good flat-fielding/illumination correction and 
sky subtraction. The lower corners of the images show the presence of the CCD supports. These parts are not used in our analysis, but are shown here for the sake of completeness.}
\label{GMOSexample}
\end{figure*}

Figure~\ref{GMOSexample} shows an example of our final stacked images
for the NB (left) and $g$-band (right) filters for
SDSSJ121503.13+003450.6. These images show the full field of view
(FOV) of our single field observation,
i.e. 5.5\arcmin~$\times$~5.5\arcmin.  From Figure~\ref{GMOSexample},
it is clear that our images are nearly flat in both NB and $g$-band
and lack significant large-scale residuals, indicating that our
flat-fielding and background subtraction are good, even though we
simply subtracted a constant background from each individual frame (see
\S~\ref{sec:datared}). The lower
corners of the images show the supports of the CCDs, and are thus
noisier in the final stack. These parts of the images are not used in
our analysis, but we show them here for completeness.
Note that 
the noise at the position of the two vertical CCD gaps ($\sim2.8$ arcsec each)
is slightly
higher. This is more evident in the NB stack compared to the $g$-band
image.  Indeed, given the typical number of only 6 dithered NB images
of 1200s, the gap has been filled, but results in slightly shallower
data for a small fraction of the image\footnote{Note that the targeted quasars are $\sim70$\arcsec 
($\sim574$ kpc) away from the CCD gaps, i.e. they are close to the center of
the GMOS-S field-of-view.}. 

Interestingly, in the NB data presented in Figure~\ref{GMOSexample} a
LAB has been identified at 69\arcsec\ from the quasar with an average
${\rm SB}_{\rm Ly\alpha}\sim10^{-17}$~\unitcgssb, with a size of
7.8\arcsec\ ($\sim 64$~kpc).  This source confirms that we are able to
detect extended emission at the level planned for our observations, at the 
characteristic SB of the giant Ly$\alpha$ nebulae around UM~287 and SDSSJ0841+3921.

As we are interested in the \lya line emission of the gas distribution
around the QSO, one might naively think to compute a continuum
subtracted image, e.g. as performed in
\citet{Cantalupo2014}. However, continuum subtraction would
inevitably increase the statistical and systematic error budget
(e.g. PSF matching, diffuse light, large galaxies in the $g$-band, unknown 
spectral slope of the stars/galaxies, and
the use of two images instead of only one),
significantly decreasing
our ability to detect the expected weak diffuse \lya emission signal.
Furthermore, in the halo of a QSO we do not expect to detect extended diffuse continuum
emission at a level comparable to the diffuse \lya emission\footnote{Note that 
the two-photon continuum expected from the same gas emitting the extended \lya emission is of the order 
of $f_\lambda \sim 10^{-20}$~erg~s$^{-1}$~cm$^{-2}$~\AA$^{-1}$ (see Figure 9 of \citealt{FAB2015}), and thus negligible in 
comparison to the \lya emission itself.}.

For these reasons, we decided to avoid the continuum-subtraction step,
and instead mask all the continuum sources detected in the $g$-band. To
build these masks we run {\tt SExtractor} on the $g$-band images to
identify all the sources down to a very low detection threshold ({\tt
  DETECT\_THRESH}$=$1.0), and allowing the detection of very compact
objects ({\tt DETECT\_MINAREA}$=$5 pixels). We then use the
`segmentation'\footnote{The `segmentation' image is already a mask of
  the image, in which each source is represented by its total
  isophotal area with flux equal to the identification number in the
  {\tt SExtractor} catalogue.}  image produced by {\tt SExtractor} to
create the final object mask for each NB stack. Given that {\tt SExtractor}
unambiguously assigns an identification number to each source in the
field, we can `switch off' the mask for the sources we are interested
in, i.e. the QSO or the stars used for the PSF comparison in our
analysis.

Furthermore, to ensure that we do not mistakenly detect \lya signal
from compact objects 
in proximity of the QSOs, i.e. the LAEs which do not have a continuum detection, we produce an analogous
mask using the NB image. However, this mask targets only compact
sources but does not mask diffuse emission, i.e. {\tt DETECT\_MINAREA}=5
pixels and {\tt DETECT\_MAXAREA}=15 pixels.  Note that this NB masking step
removes only a small fraction of pixels from our analysis,
i.e. $\sim1.7$\% from the whole image, or $\sim1.6$\% within the
1\arcmin$\times$1\arcmin\ FOV centered on the quasar, with little
variation from field to field.
To generate the masks, we do not
convolve the images with a spatial filter in either of the {\tt
  SExtractor} runs.

We then generate a final mask by combining the two individual masks, so that both pixels from the $g$-band and NB masks are neglected in our analysis.
After this masking procedure has been adopted, the fraction of the FOV usable around each quasar on average is about $84$\%,  while within a region of 1\arcmin$\times$1\arcmin centered on the
quasar it is about $87$\%, with obvious variations from field to field.

\begin{figure}
\centering
\epsfig{file=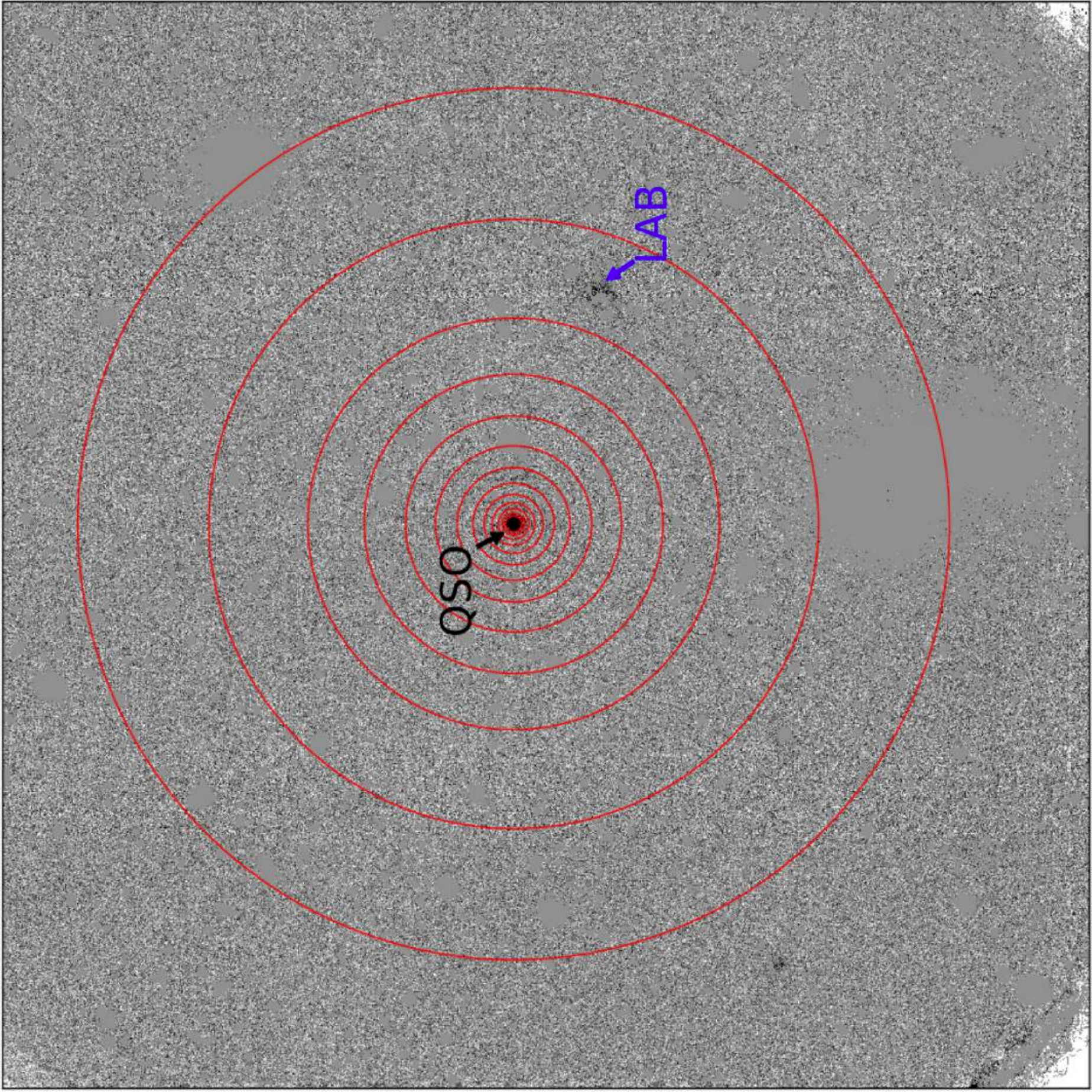, width=0.49\textwidth, angle=270, clip} 
\caption[Example of a NB image after masking all the sources.]{NB image after masking all the sources as explained in the text. 
The masked sources are set to zero (gray color, see Figure~\ref{GMOS_all} for a better visualization of all the masks).
This image shows the field around the quasar SDSSJ121503.13+003450.6, as in Fig.~\ref{GMOSexample}, for comparison. 
Note in particular that the diffuse emission of the LAB is only masked where compact continuum sources are identified, i.e. the LAB is still visible.
For reference, we overplot in red the annuli within which we calculate radial profiles in our analysis.}
\label{MaskExample}
\end{figure}

Figure~\ref{MaskExample} shows the NB image of SDSSJ121503.13+003450.6
this time after masking all the sources except the QSO, using the final
mask. It is clear that our method does not mask extended emission in
the NB images. Indeed, although there are compact continuum sources within the
LAB detected in the NB image, the extended emission is still clearly visible after applying the final mask (compare
Fig.~\ref{GMOSexample} with Fig.~\ref{MaskExample}).
For reference,
in this figure we overplot in red the annuli within which we calculate
the radial SB profiles in the next sections.

\subsection{Individual NB Images vs Giant \lya Nebulae and EELR}
\label{PSF}

\begin{figure*}
\centering
\epsfig{file=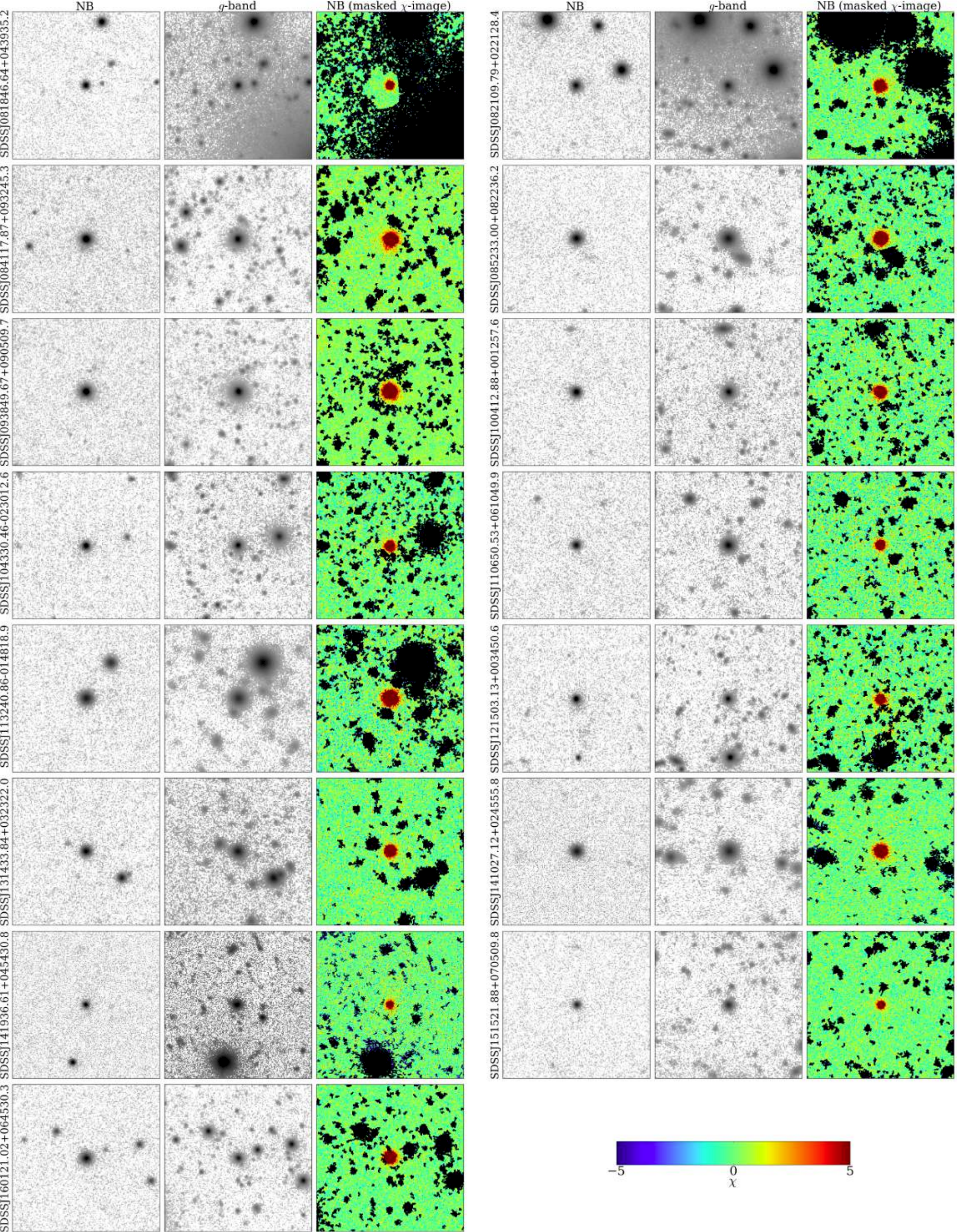, width=1.01\textwidth, clip} 
\caption{Postage-stamp images of 1\arcmin$\times$1\arcmin\ (corresponding to about $492$~kpc $\times492$~kpc at $z=2.253$) centered on each quasar of our sample.
For each quasar we show from left to right: the NB image (log-scale), the $g$-band image (log-scale), and the $\chi$-image (linear-scale) with superimposed the final mask (black).
Note the absence of extended \lya emission on scales of $\sim100$~kpc ($\sim12$\arcsec) as opposed to the observations of 
UM~287 (\citealt{Cantalupo2014}) or SDSSJ0841+3921 (\citealt{Jackpot}), even though we should 
have been able to detect such emission. North is up, East is left.}
\label{GMOS_all}
\end{figure*}

In Figure~\ref{GMOS_all} we show images in the NB filter and
$g$-band for all the 15 quasars observed with GMOS-S in the
FLASHLIGHT survey. We also show $\chi$-images in the NB filter with
the final mask for each field (black) superposed.  The $\chi$ images
are computed as $\chi=I/\sigma$, where $I$ is the final stack and
$\sigma$ is the square root of the variance image.  Indeed, as
previously described,
our custom pipeline results in a
final stack for each filter (NB and $g$-band), and a final variance
image ($\sigma^2$), which contains all the information on the
error budget. In the $\chi$ images, emission will be manifest as
residual flux, inconsistent with being Gaussian distributed noise (if
the noise has been correctly propagated, and thus the $\sigma^2$ image
is an accurate description of the noise in the data).
From these images it is
clear that we have not detected any giant \lya nebulae similar to
UM~287 (\citealt{Cantalupo2014}) or SDSSJ0841+3921
(\citealt{Jackpot}), i.e.
SB$_{\rm Ly\alpha}\sim 10^{-17}$ \unitcgssb on scales $> 50$~kpc,
even though we should have
been able to easily detect such giant \lya nebulae given the SB limits quoted in Table~1. 

Furthermore, we have also searched our data for emission on smaller scales, 
i.e. the emission typical of EELR  (e.g., \citealt{Husemann2013, Husemann2014}) or so called ``\lya fuzz'' (e.g., \citealt{Heckman1991,Christensen2006, Hennawi2013, Herenz2015}), 
by subtracting a crude estimate of the unresolved emission of each quasar. We proceed as follows.
First, as we are interested in the \lya line emission, we used the $g$-band images to evaluate the continuum 
emission of the quasar and subtract it from the NB images, following 
\citet{Yang2009} and \citet{FAB2015LAB}, as the \lya line is within the $g$-band filter (but close to the edge, at low transmission, i.e. $35$\%):

\begin{equation}
f_{\lambda,\, {\rm cont}}^{g} = \frac{F_{g} - F_{\rm N\!B}}{\Delta\lambda_{g} - \Delta\lambda_{\rm N\!B} }
\end{equation} 

\begin{equation}
F_{\rm Ly\alpha} = F_{\rm N\!B} - a f_{\lambda,\, {\rm cont}}^{g}\Delta\lambda_{\rm N\!B},
\end{equation} 
where $F_{g}$ is the flux in the $g$-band, $F_{\rm N\!B}$ is
the flux in the NB filter, $\Delta\lambda_{g}$ and
$\Delta\lambda_{\rm N\!B}$ represent the FWHM of the $g$-band and NB
filter, respectively, $f_{\lambda,\, {\rm cont}}^{g}$ is the
flux density of the continuum within the $g$-band, and $F_{\rm
Ly\alpha}$ is the \lya line flux. The parameter $a$ allows a better match to the continuum, 
and ranges between 0.6 and 1.0 in our sample, reflecting our ignorance on the slope of the continuum for individual sources, as probed by the $g$ and NB filters.
Note that this range of value is in agreement with the expected continuum slope of quasars in this wavelength range (\citealt{VandenBerk2001}).

After the continuum subtraction, the images contain information only on the Ly$\alpha$ emission, 
but still contaminated from the unresolved line emission of the quasar. 
To removed the unresolved Ly$\alpha$ emission, 
we have assumed that the $g$-band
image of the quasar represents a good estimate of the
point-spread-function (PSF) of our observations\footnote{As the seeing
  in the $g$-band and NB are in agreement, this is true for both
  $g$-band and narrow-band (see section \ref{PSF} for more detailed
  analysis of the PSF). Note that in this analysis we used the masked images.}.
We have then rescaled the peak of this PSF model to
the peak of the quasar \lya image, and subtracted it.  In this way, we
are left with a map of the \lya line which is cleaned from the
unresolved line emission emerging from the broad-line regions (BLR; e.g,
\citealt{Stern2015}) of the quasar, and the unresolved continuum emission, e.g. from the accretion disk or host galaxy.

\begin{figure*}
\centering
\epsfig{file=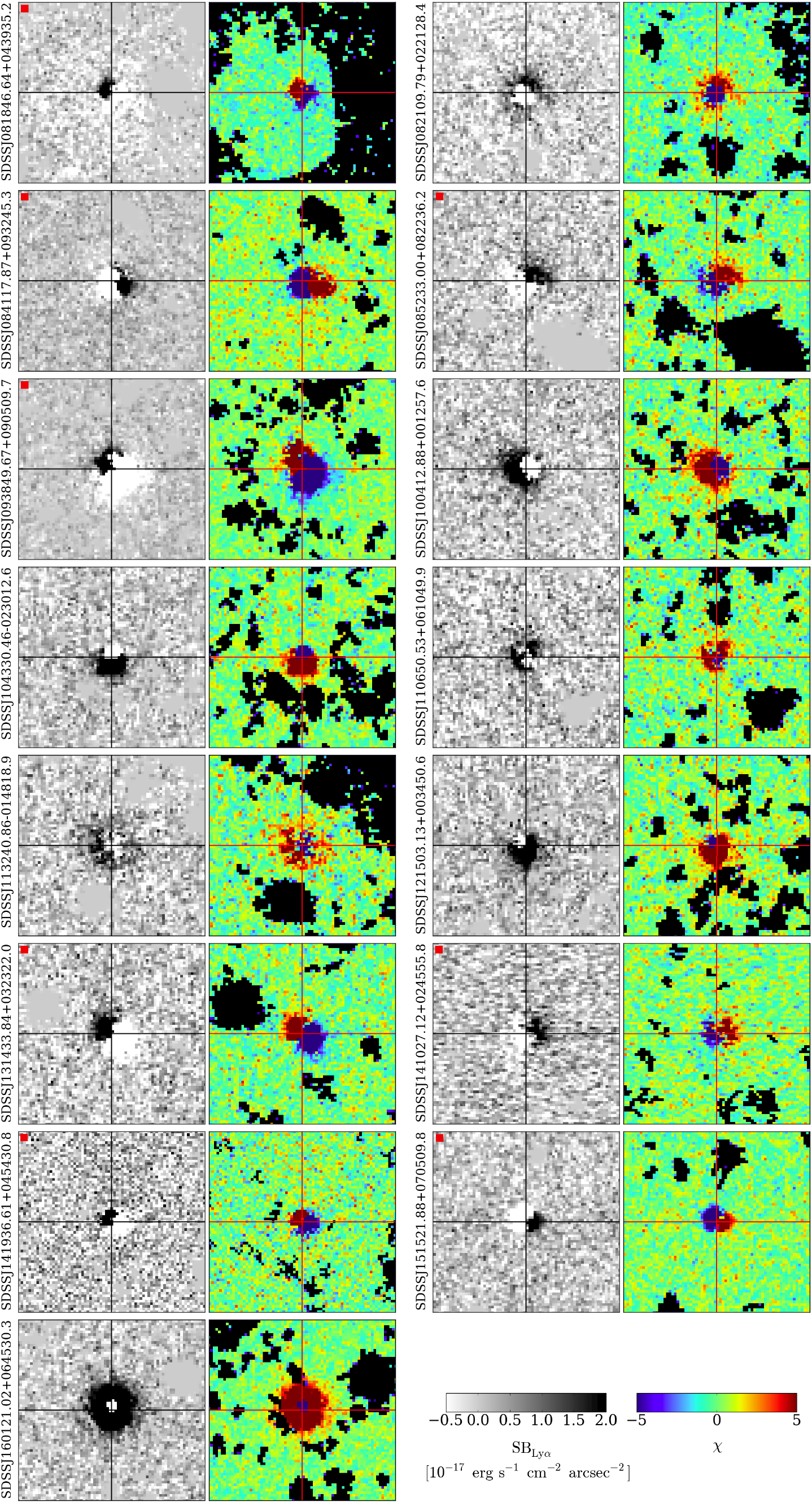, width=0.7\textwidth, clip} 
\caption{\lya surface brightness maps (linear-scale) and $\chi$-images (linear-scale) of 20\arcsec$\times$20\arcsec\ (corresponding to about $164$~kpc $\times164$~kpc at $z=2.253$) 
after continuum and PSF subtraction centered on each quasar of our sample.
The position of the peak of the quasar before PSF subtraction is indicated by the cross.
The PSF subtraction reveals \lya emission above SB$_{{\rm Ly}\alpha} = 10^{-17}$ \unitcgssb extending on scales of radius $\sim 3-4$ 
arcsec ($\sim 25-32$~kpc) around 7 out of 15 quasars, i.e. 47\% of the sample. The quasars whose PSF subtraction reveal stronger residuals are indicated with a red square in the top-left corner of their image.
North is up, East is to the left.
}
\label{Fig4}
\end{figure*}

In Figure~\ref{Fig4} we show the results of this attempt. For each
quasar we present a 20\arcsec$\times$20\arcsec\ \lya maps in surface
brightness units, and the correspective $\chi$-images, with both
images masked using the final mask computed using the aforementioned
method.  The PSF subtraction reveals \lya emission above SB$_{{\rm
    Ly}\alpha} = 10^{-17}$ \unitcgssb extending on scales of radius
$\sim 3-4$ arcsec ($\sim 25-32$~kpc) around 7 out of 15 quasars,
i.e. $47$~\% of the sample, while the other objects show more compact
residuals, or residuals affected by systematics in the PSF subtraction.
Indeed, the other 8 sources (indicated with a red square in the top-left corner) show asymmetrical residuals, with positive and negative fluctuations clearly visible in 
the $\chi$ images presented in Figure~\ref{Fig4}.
These residuals could be due both to systematics introduced by the use of the $g$-band of the quasar as PSF, 
and/or to intrinsic asymmetries in the emission on these small scales. 
A careful PSF subtraction to study the emission 
on $\lesssim$50~kpc scale is beyond the scope of this work, but we stress that would be more accurate 
in the case of integral field spectroscopy (e.g., \citealt{Husemann2014}).   
Note that, probably because of these systematics in the PSF subtraction, our detection rate for these `\lya fuzz' is at the low end of the 
frequency range of $\sim 50 -70$\% deduced in the literature (\citealt{Christensen2006, Courbin2008, Hennawi2009, North2012}), 
but higher, as expected, than long-slit spectroscopic searches, i.e. $34$\% (\citealt{Hennawi2013}).

In summary, the analysis of the individual objects do {\it not}
present any giant \lya nebula similar to UM~287
(\citealt{Cantalupo2014}) or SDSSJ0841+3921 (\citealt{Jackpot}), but
only \lya emission on scales $< 50$~kpc. We ascribe this emission to
the EELR or the so called ``\lya fuzz'', or in other words, to gas photoionized by the central AGN or 
by star-formation in the host galaxy of the quasar (\citealt{Heckman1991, Husemann2014}). Other mechanisms could also contribute, e.g. 
shock-heated gas, scattered \lya emission. In the subsequent sections
we perform the stacking analysis to search for the signal from the
quasar CGM, i.e. on larger scales.

\subsection{Constraining the PSF of the NB Images}
\label{PSF}

To quantify the extended emission in the \lya line around the quasars
in our fields, i.e. not related to unresolved emission from the quasar
itself, we have to carefully estimate the PSF of our NB images.  We
need to compute the PSF with high accuracy, in order to detect even
small deviations from its shape in the average profile of the QSOs.
For this purpose, we match the SDSS star catalogue with the sources in
our 15 NB fields and select all the high S/N stars. This operation
resulted in a sample of 115 usable stars. For each of these stars, we
have created the NB masked image as explained in \S\ref{imPrep}, and
calculated the stars emission
profile in radial logarithmic bins (see red circles in
Figure~\ref{MaskExample}) with an unweighted average of all the
not-masked pixels within each annulus. We then consistently propagate 
the errors from the
variance image\footnote{In the remainder of the
  paper we use the same method for the profile extraction.}.  The
profiles of the 115 stars are
then averaged with equal weights
to obtain the PSF
of the NB images\footnote{Note that we use the maximum number of available stars to minimize the errors, even if this led to a different number of stars per image. 
This effect could in principle bias the PSF characterization towards an image. However, we have found 5-10 usable stars per field, with only two fields without
available stars for our analysis. This sampling should not affect significantly our results given that the data were taken under similar weather conditions.}.

\begin{figure}[h!]
\centering
\epsfig{file=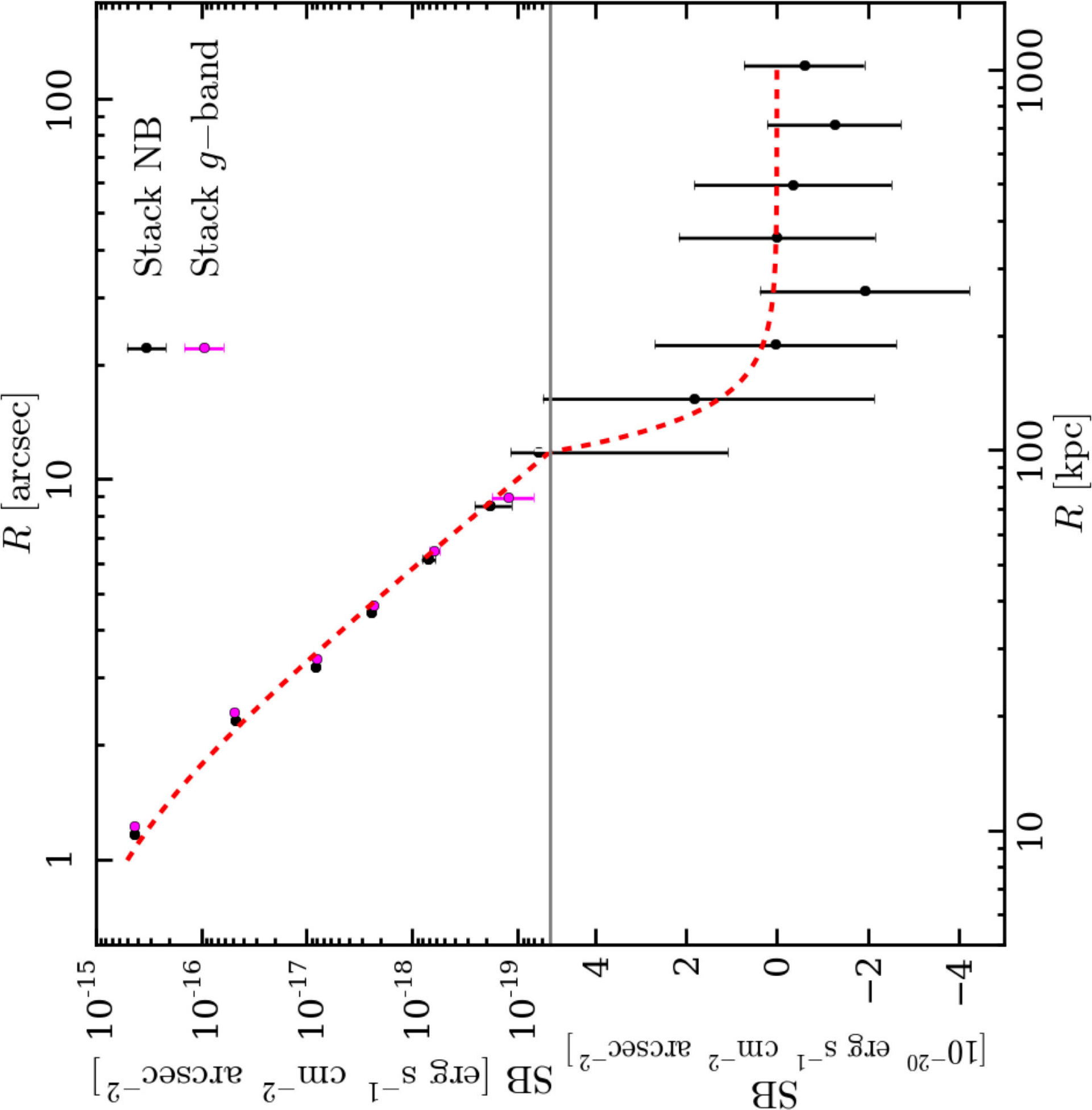, width=1.0\columnwidth, angle=270, clip} 
\caption[PSF determination: average radial profile of 115 stars in our NB images.]{Average
  radial profile for 115 stars in our NB images (black points). 
This PSF is well fitted by a Moffat profile with FWHM$=1.2$\arcsec, and $\beta=2.15$ (red dashed line). Note the different scale in the plot, i.e. log-log plot to 
capture the whole range of the PSF, and log-linear to check the zero level at large radii. The PSF of the NB images (black) is in good agreement with the PSF of the $g$-band (magenta). 
Note that the $g$-band data points have been normalized to the NB observations (see text for details), and artificially moved at slightly higher radii to avoid superposition with the NB data points.}
\label{starsPSF}
\end{figure}

Figure~\ref{starsPSF} shows the average 
radial profile for
the 115 high S/N stars in our NB images (black points). The profile is
consistent with a Moffat function, as expected for seeing limited
observations, where the PSF is determined by the wiggling of the stars
on the CCD (\citealt{Trujillo2001}). The normalized Moffat profile is
defined as
\begin{equation}
{\rm PSF}(r)=\frac{\beta-1}{\pi\alpha ^2}\left[1+\left(\frac{r}{\alpha}\right)^2\right]^{-\beta},
\label{Moffat}
\end{equation}
where the full width at half maximum is given by
FWHM$=2\alpha\sqrt{(2^{1/\beta}-1)}$, and the total flux is normalized
to 1. We fit the average profile of the stars with this function and
obtained best fit parameter values of FWHM=1.2\arcsec and
$\beta=2.15$. The FWHM value we obtained is perfectly consistent with
the independent measurement of the median seeing in our observations
(see \S\ref{sec:data}) estimated by selecting all the stars in the
images, and applying the {\tt psfmeasure} task within the IRAF
software package.  In the absence of imperfections in telescope
optics, the $\beta$ parameter in eqn.~(\ref{Moffat}) should have a
value of $\sim4.765$, as expected from turbulence theory (e.g.,
\citealt{Saglia1993, Trujillo2001}). However, it is known that PSFs
typically measured in real images have larger ``wings'', or
equivalently, smaller values of $\beta$. This is due to the fact that
the real seeing also depends on the performance of the telescope
optics, and not only on the atmospheric conditions
(\citealt{Trujillo2001}). Our value is in agreement with this picture,
however there is no tabulated PSF for the GMOS-S instrument (German
Gimeno\footnote{German Gimeno is the current Instrument Scientist for
  GMOS-S.}, private communication).

To test our PSF estimate, we perform the same calculation on masked $g$-band images. 
We select 15 high S/N not saturated stars, i.e.
one per field to include in our calculation all possible
seeing variations between different stacks. We then calculate the average radial profile, as for the stars in the NB images. Figure~\ref{starsPSF} shows (in magenta) 
this average radial profile after normalizing it to the NB data points. The normalization factor is computed as the ratio between the peak values of the two average profiles. 
The shape of the Moffat profile of the NB and $g$-band images are in remarkable agreement.
Note that in the case of the $g$-band PSF calculation, we probe only smaller radii,
due to confusion arising mainly from diffuse light from bright objects.
Finally, note that the Moffat profile goes smoothly to zero at large radii, and the average 
star profile is consistent with this within our uncertainties 
(see log-linear region of the plot in Fig.~\ref{starsPSF}).

Having characterized the PSF of our observations, we can now ascertain whether our sample of QSOs exhibit a signal from the surrounding diffuse gas distribution.  
To do so, we apply the same stacking analysis to the quasar sample.

\section{Results}
\label{sec:results}

\begin{figure}
\centering
\epsfig{file=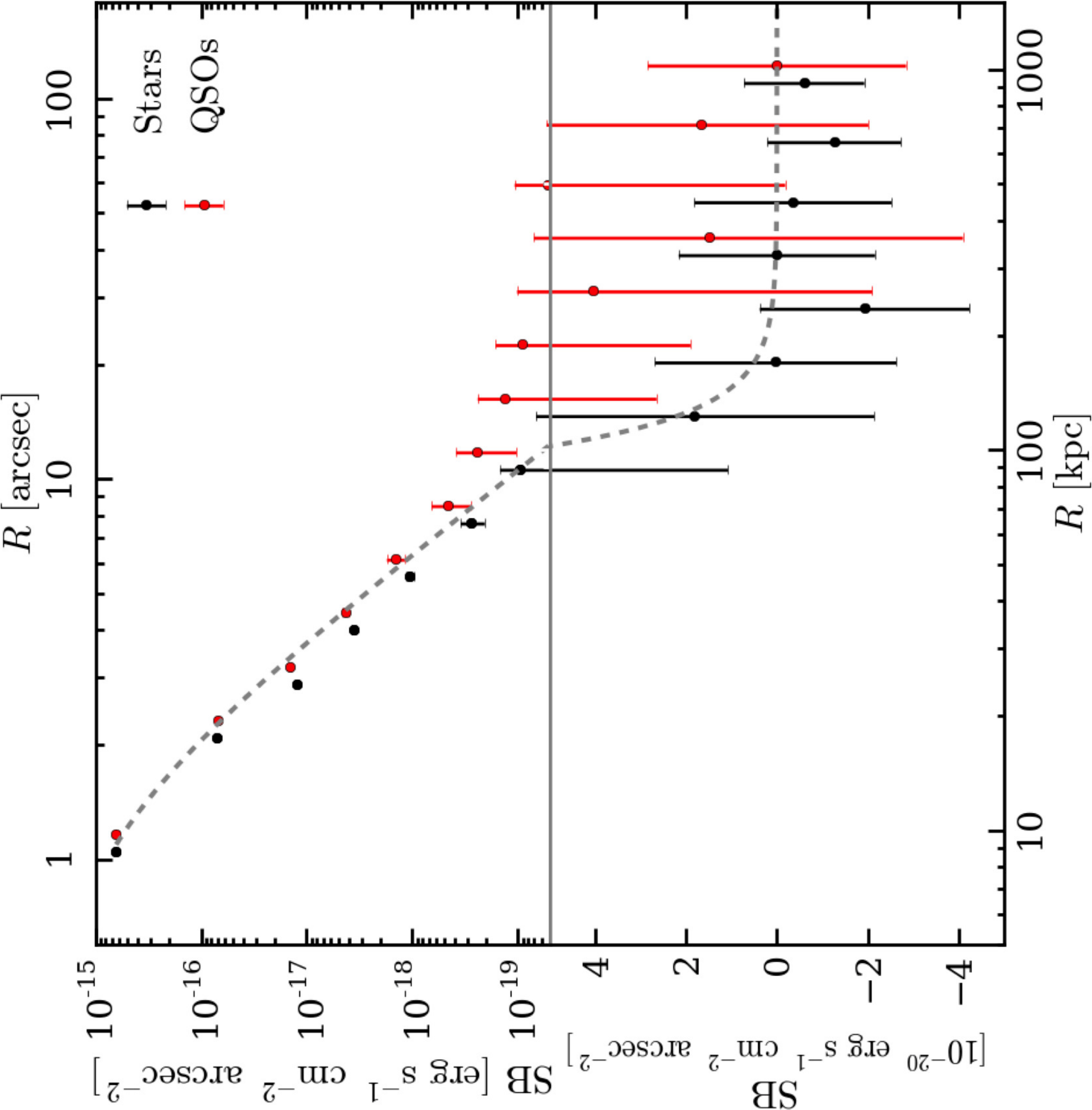, width=0.98\columnwidth, angle=270, clip} 
\caption[Comparison between the stacked QSO profile and the PSF of the NB images.]{Average radial profile of 15 QSOs compared to the average star profile normalized to the QSO data points. 
The plot is divided into a log-log plot to 
show the whole range of the profiles, and a lin-log plot showing that the QSOs' profile is also consistent with zero at large radii. 
The gray dashed line indicate the Moffat profile with parameters derived from the fit to the stars' profile.
The QSOs' profile shows a deviation from the stars' 
PSF at SB~$\sim10^{-19}$~erg\,s$^{-1}$\,cm$^{-2}$\,arcsec$^{-2}$. 
The stars' profile is slightly shifted to smaller radii 
for clarity (e.g. to avoid the superposition of the errorbars).}
\label{comp_profiles_cool}
\end{figure}

As done for the stars, we have computed the radial profile for each
quasar from the NB masked images. The 15 quasar profiles are then
averaged to obtain the average signal around $z=2.253$
quasars.  Figure~\ref{comp_profiles_cool} shows the average 
radial profile of the QSOs, together with the average profile
of the stars normalized to the QSOs data points.
The normalization has been performed by multiplying the average star profile 
by the ratio between the peak values of the two average profiles, i.e. the first bin.
As the first annulus for the profile extraction has a radius of 1.2\arcsec\, this is also consistent with an approximate normalization in flux.
In addition, for comparison purposes, in this plot we slightly shift
the stars' PSF profile to smaller radii,
i.e. the profiles are estimated with the same radial bins, but are
shown at different locations to avoid the overlap of the error-bars.
The QSOs' profile clearly deviates from a pure Moffat function at
large radii ($\gtrsim70$~kpc), and at a low surface brightness level of SB$\sim10^{-19}$ \unitcgssb, 
and, as expected, is consistent with zero at very large radii ($\gtrsim600$~kpc). 

\begin{figure}
\centering
\epsfig{file=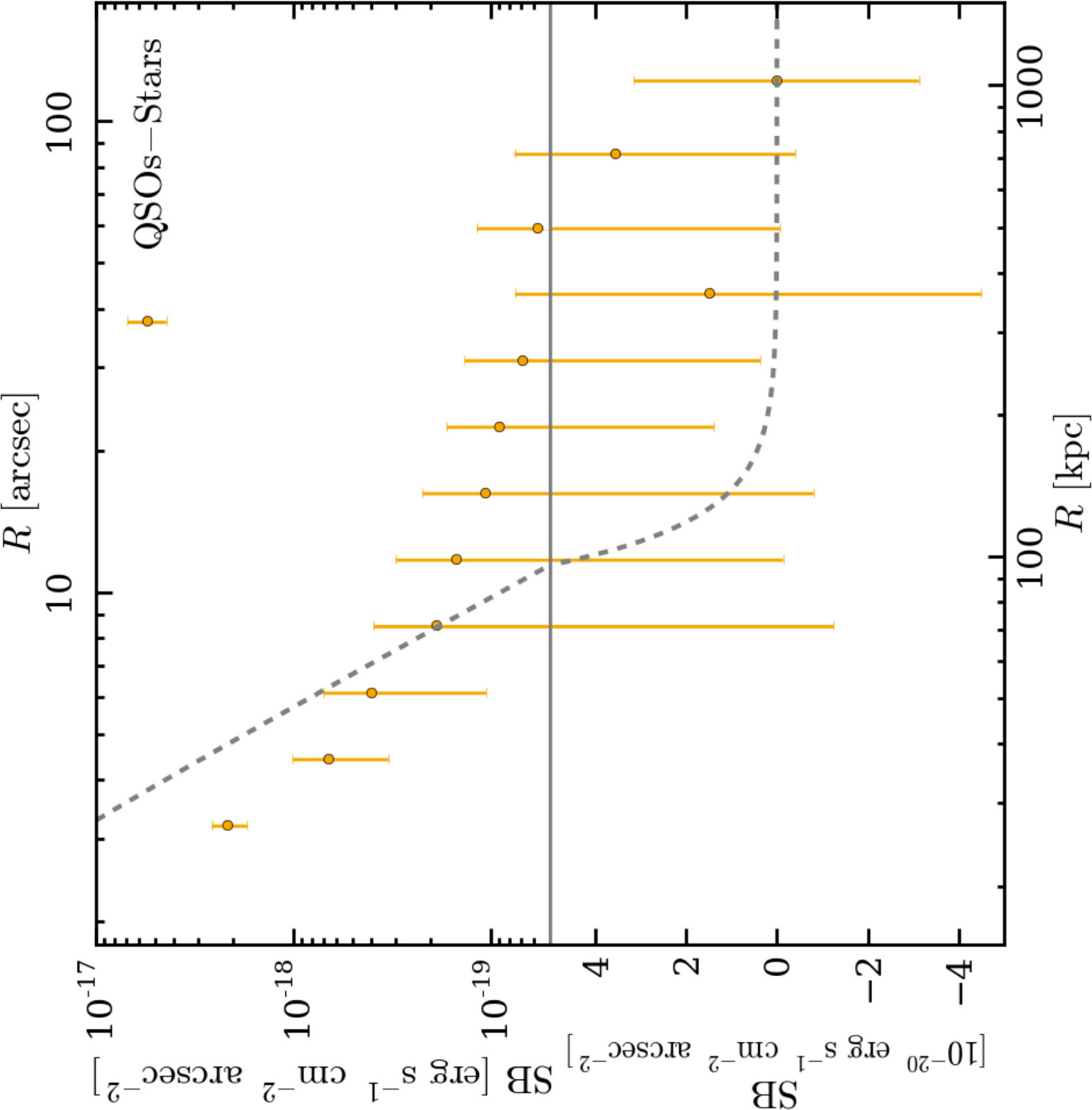, width=0.98\columnwidth, angle=270, clip} 
\caption{
Difference of the average QSO profile and the normalized stars profile. This may be interpreted as the average SB$_{\rm Ly\alpha}$ profile 
from the gas distribution around a typical $z\sim2$ QSO. Note the extremely low surface brightness probed at large distances from the quasars.}
\label{comp_profiles_norm}
\end{figure}

This extended emission is more evident if we take the difference between the QSOs' profile and the normalized stars' profile, 
or in other words, if we remove the contribution from the unresolved emission of the QSOs.
The resulting profile thus represents the average \lya emission arising from the distribution of the cool gas (${T\sim10^4}$~K) around a typical bright $z\sim2$ QSO. 
Figure~\ref{comp_profiles_norm} displays the profile of the difference
between QSOs and the normalized star profile, which extends for
hundreds of kpc to very low SB levels SB$_{\rm
  Ly\alpha}\sim10^{-19}-10^{-20}$~erg\,s$^{-1}$\,cm$^{-2}$\,arcsec$^{-2}$.
However, we stress again that 
we do not subtract the continuum emission from the NB images and thus, on small scales, i.e. the scale of the quasar's host (tens of kpc), the signal we see might be due to contamination from the resolved host galaxy or is 
emitted within the aforementioned EELR.
If this is due to the host galaxy, we estimated from our profile that 
the galaxy should have $f_{\lambda}=2.03\times10^{-17}$~erg~s$^{-1}$~cm$^{-2}$~\AA$^{-1}$ within an aperture of
3 arcsec radius ($\sim25$~kpc), or equivalently a magnitude of 23.9.
This value 
is similar to the rest-frame UV apparent magnitude of a bright Lyman break galaxy (LBG) at comparable redshift (e.g., \citealt{Reddy2009}), 
and thus brighter than a $L_{\star}$ galaxy for LBGs, i.e. $m_{\star}=24.54$ ($R_{\rm AB}$; \citealt{Adelberger2000}).

In addition, we check if the signal we detect is dominated by a single bright object
in the stack by repeating the same
experiment each time removing a different QSO from our sample.
Specifically, we compute 15 average profiles of 14 QSOs, each
time removing a different QSO.  Then, to obtain the CGM profiles, we
subtract from each of them the normalized star profile. In this way,
it would be clear if a bright QSO in our sample dominates the stack at
any radius. 

\begin{figure}
\centering
\epsfig{file=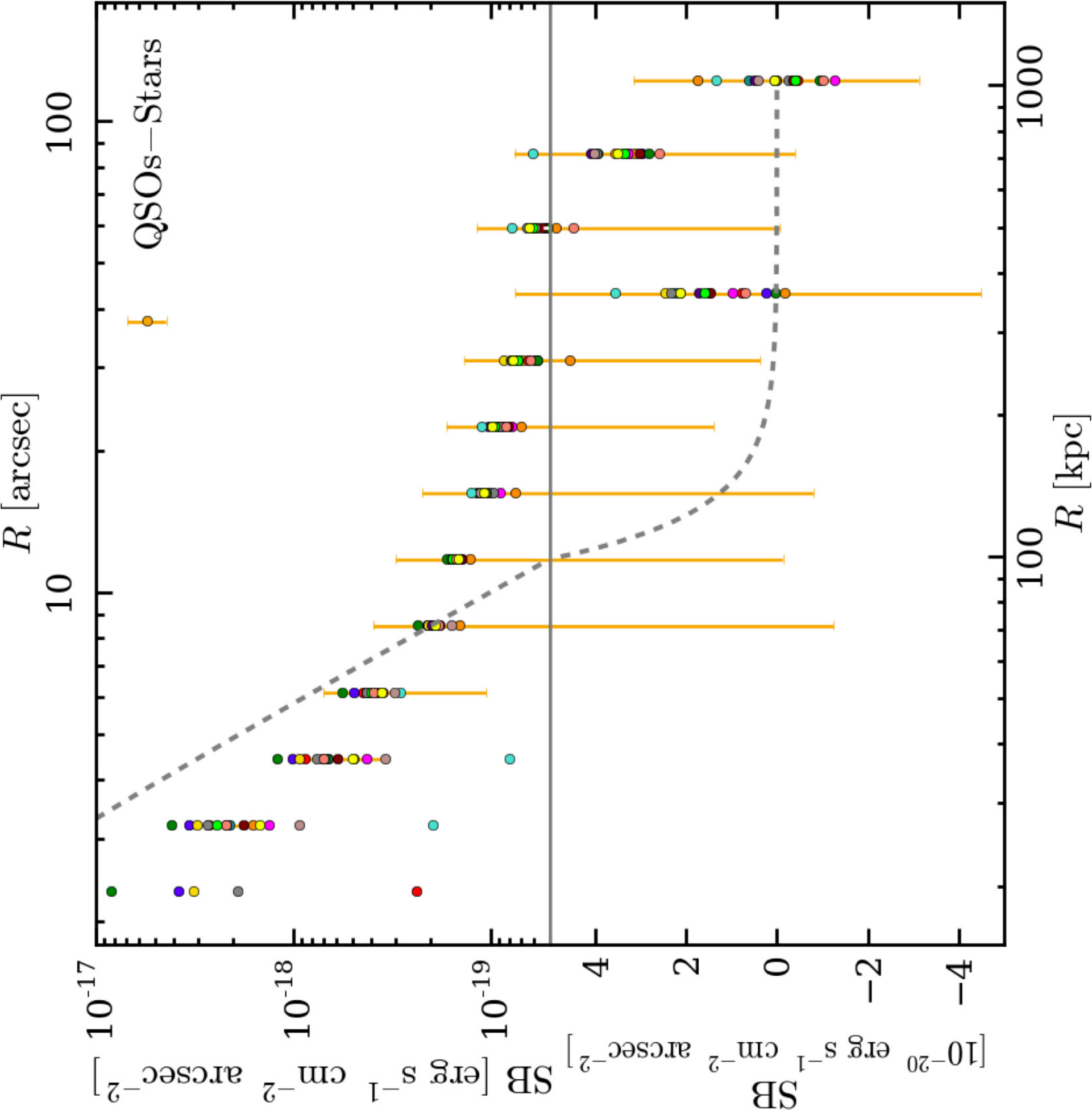, width=0.98\columnwidth, angle=270, clip} 
\caption[Dependence of the CGM profile on each individual QSO.]{Dependence of the \lya profile on each individual QSO. We estimate the difference between the QSOs' profile and the normalized
stars' profile 15 times, each time by removing a different QSO from the QSOs' profile calculation. The different profiles are shown with different colors. 
It is clear, that none of the QSOs of the sample is dominating the \lya profile for $R>50$~kpc.
Larger variations, i.e. larger than the $1\sigma$ error, occur at smaller radii. This is probably due to slightly different seeing variations or host galaxy contamination (see \S\ref{sec:results}).}
\label{QSOrel}
\end{figure}

Figure~\ref{QSOrel} shows the results of this test. It is clear that
the profile is not changing significantly between different
calculations for $R>50$~kpc. For smaller radii, as explained above,
differences could arise due to seeing variations or host galaxy
contamination. However, this is not the focus of the current work, and
we are confident that none of the QSOs in our sample dominates the
average profile on the large scales $R>50$~kpc characteristic of the CGM. 
However, given that the signal is close to our detection limit 
we verify the significance of our results in the next section. 

\subsection{Testing the Significance of the Results with a Monte Carlo Analysis}
\label{sec:Monte Carlo}

\begin{figure*}
\centering
\epsfig{file=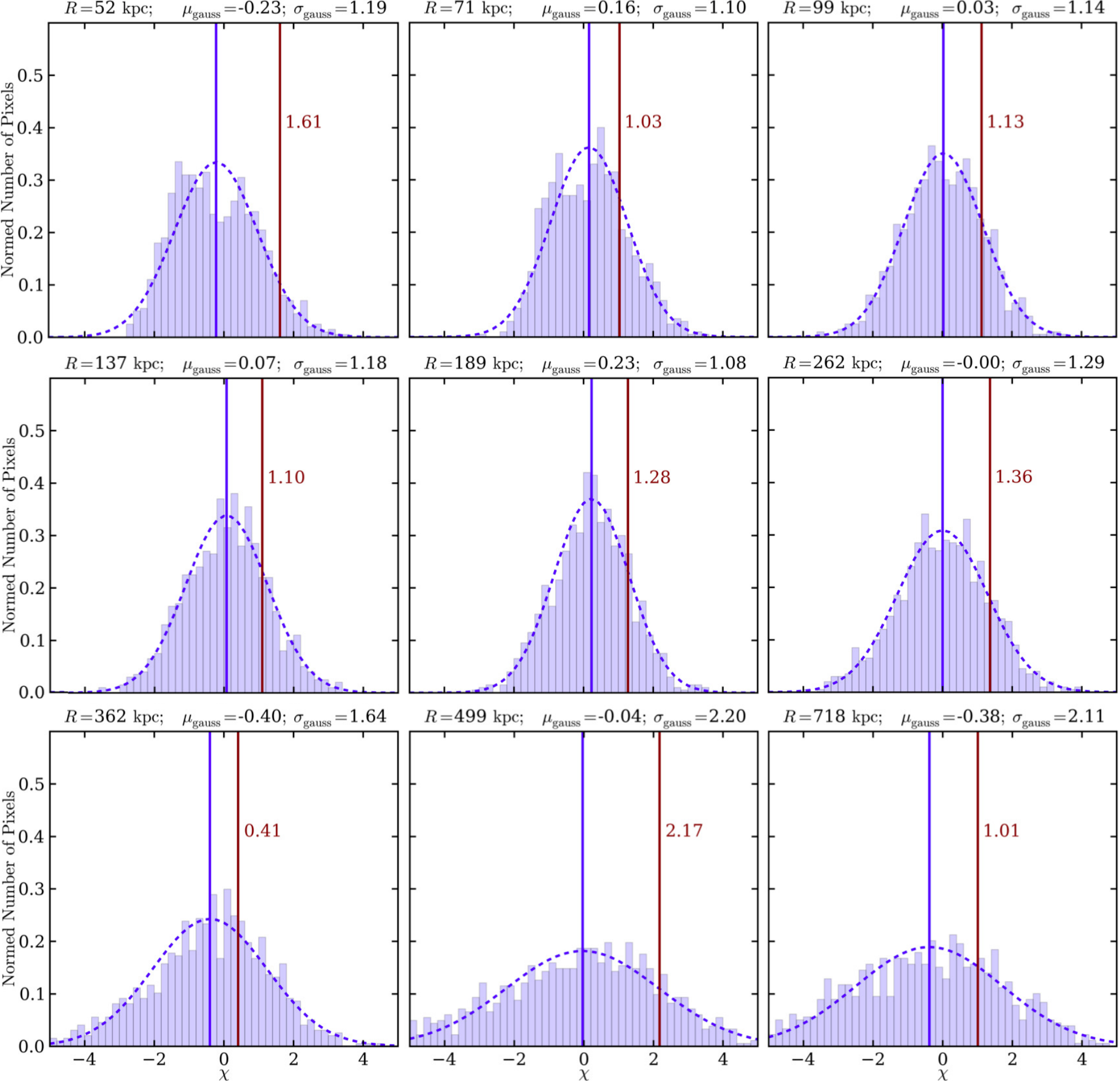, width=0.98\textwidth} 
\caption[$\chi$ histograms of 1000 realizations of our same experiment (Monte Carlo analysis).]{$\chi$ histograms of the 1000 realizations of our same experiment (see Section for details) using random samples of 15 stars each. We show the histograms for each bin starting from $\sim50$~kpc.
The histograms are consistent with Gaussian distributions centered in zero (see the central $\mu_{\rm gauss}$ values) and with nearly unit variance (see the $\sigma_{\rm gauss}$ values) till $\sim300$~kpc. 
The bins at larger radii seem to deviate from the simple Poisson statistics. The blue vertical lines indicate the central value of each histogram ($\mu_{\rm gauss}$), 
while the vertical brown lines indicate the $\chi$ for our measurements in Fig.~\ref{comp_profiles_norm}.}
\label{MCMC_1}
\end{figure*}

Given the faintness of the emission we are looking for, we have applied
an {\it ad-hoc} complicated preparation of our images before the profile
extraction. Although this effort suggests a tentative detection, our
results could still be dominated by systematic errors. In particular,
our measurement of the CGM radial profile out to very large radii
could be affected by 1) errors in the empirical PSF subtraction of the
quasar profile using the PSF of stars, including errors in the
re-scaling of the stellar PSF; and 2) systematics in the data reduction and in
the data itself (e.g., flat-fielding, constant sky-subtraction,
ccd-noise, masking). These systematics could prevent the noise from
averaging down according to simple photon-counting statistics, i.e.
the expected error on the average SB should be inversely proportional to the
square root of the area employed in the average ${\sigma_{\rm SB}^{\rm Area}=\sigma_{\rm SB}^{\rm 1
    arcsec^2}/\sqrt{\rm Area}}$. 

For these reasons, to firmly assess the statistical significance of our results,
we perform a Monte Carlo analysis. We have built $1000$ samples of 15
stars, by randomly selecting objects from the 115 stars that were used
to determine the PSF of the NB images (see \S\ref{PSF}). For each
sample of stars we have then computed the average radial
profile, and subtracted from it the normalized average profile of the
global sample of the stars. In this way, we are left with $1000$
realizations of our same experiment, i.e. QSOs' profile~$-$~stars'
profile.  These profiles should be consistent with zero at each
radius, and the $\chi$ values, i.e. the ratio between the signal and the 1$\sigma$ error within each bin,
should be consistent with a Gaussian distribution of unit variance.  If however
we find Gaussian distributions with variances greater
than 1, this would give us information on the degree to which we are underestimating the
noise, and/or at which radius systematics begin to dominate the error budget.
Furthermore, these Monte Carlo simulations of the null experiment will allow us to robustly
estimate the statistical significance of our detection in a way that also folds in both the formal
statistical errors, as well as the systematic errors.

\begin{figure}
\centering
\epsfig{file=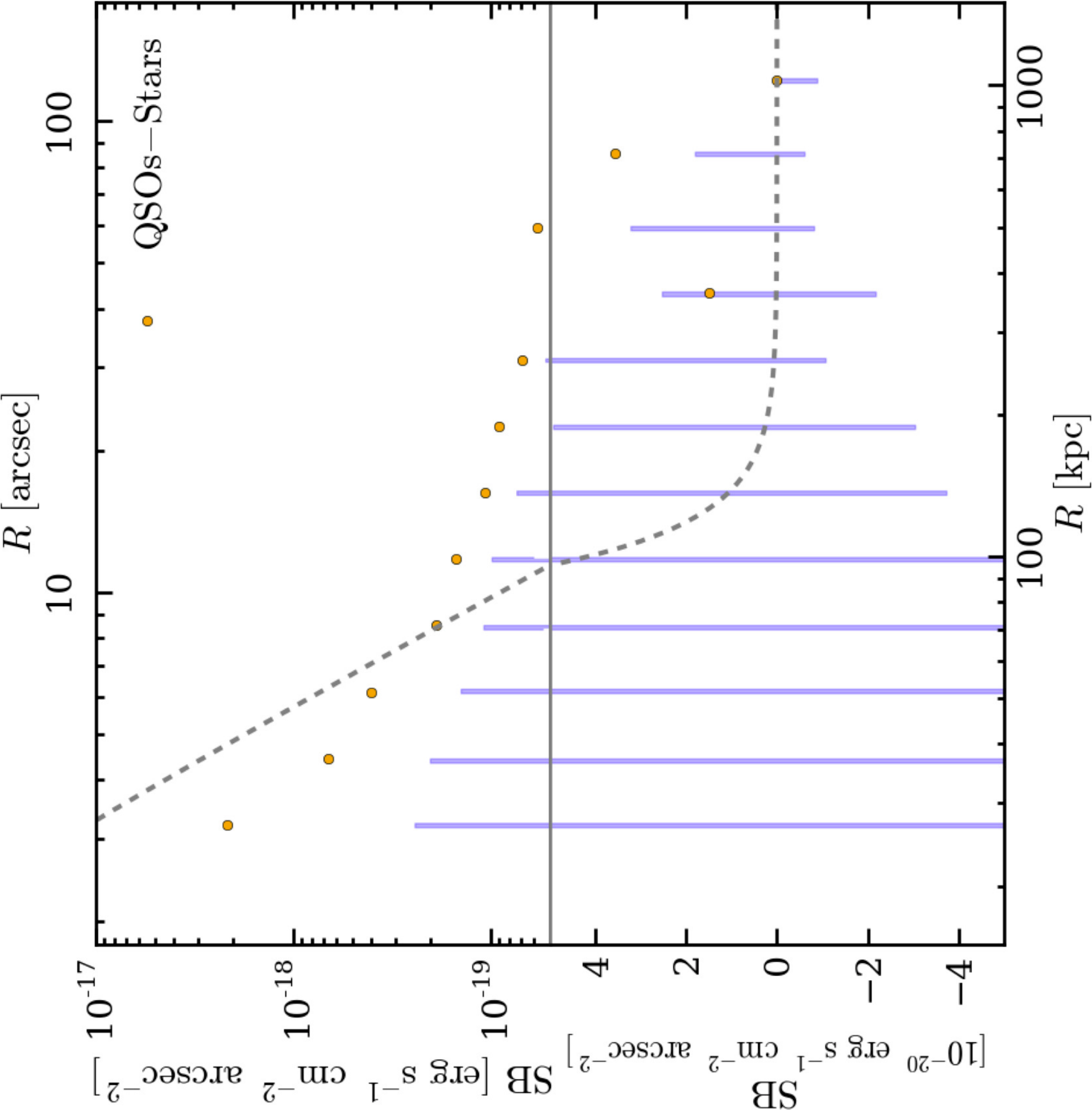, width=0.98\columnwidth, angle=270, clip} 
\caption[Statistical significance of the CGM radial profile.]{
The difference between the QSOs' profile and the average profile of the 115 stars (yellow) is compared with the confidence 
area within the 16th and the 84th percentiles from the Monte Carlo analysis (blue shaded area).
Using the current radial bins, there is only a hint for \lya signal, as already clear from Fig.~\ref{MCMC_1}.}
\label{MCMC_2}
\end{figure}

In Figure~\ref{MCMC_1} we show the distribution of the $\chi$ values for these 1000 realizations in each radial bin, 
starting from the $\sim50$~kpc radius, and compare them to our results. 
We focus on the histograms for the larger radii because, as discussed in the previous section, 
the smaller scales can be contaminated by EELR or host galaxy emission which
are not the focus of the current study. 
It is clear that the histograms have an approximately Gaussian shape,
well centered on zero (blue vertical lines). However the variances
increase at large radii, indicating that the error budget is deviating
from the expectation based on photon counting statistics.  Indeed, for ${R\lesssim200}$~kpc, we only
slightly underestimate the noise (by $\sim15$\%), while at larger radii
we incur significantly larger errors
than photon counting would suggest.
This is likely due to the fact that bins at larger radii, i.e. $\gtrsim400$~kpc, probe larger area and the implied SB limit
is significantly lower. As our SB limit decreases we become more sensitive to subtle systematics
in the data, e.g. correlations between neighboring pixels, which are larger than the formal photon counting error, and which may not average down
as $\sqrt{\rm Area}$. 
As we noted at the
beginning of this section, this deviation can be ascribed to several
possible effects.

The brown vertical lines
in Figure~\ref{MCMC_1} indicate the measurements obtained in the previous section for comparison (Figure~\ref{comp_profiles_norm}). 
Note that all our measurements constitute a positive fluctuation
which is not expected for a spurious signal\footnote{This statement
  assumes that the measurements in each radial bin are independent,
  which is a reasonable assumption provided that there is not a large
  scale systematic that correlates multiple bins.}.  In other words,
the Monte Carlo realizations are well centered around zero, while our
measurements always exhibit positive values.  Given the current
binning, with the exception of the bin
at 362 kpc, we have a $> 1\sigma$ fluctuation in every bin (or in other words above the 84th
percentile of the Monte-Carlo realizations).

This can be better visualized in Figure~\ref{MCMC_2}, where we highlight the area within the 16th and 84th percentile of the 1000 realizations in comparison to the
difference profile between QSOs and stars, e.g. same profile of Figure~\ref{comp_profiles_norm}.  
Note that in agreement with the previous histograms (Fig.~\ref{MCMC_1}), the 16th percentile of the 1000 realizations is negative. 
This can be seen in the log-lin part of Figure~\ref{MCMC_2}.
Note however that our Monte-Carlo realizations are not well centered around zero following the uncertainties expected in some of the annuli, i.e. the error on the mean should be given by
$\sigma_{\rm gauss}/\sqrt{1000}$ and thus fluctuations of our mean values about zero should be comparable to this.
In other words, we stress that the presence of systematics also appears to give rise to larger fluctuations about zero in our
Monte-Carlo realizations than simple Gaussian statistics would suggest. 
However, it is reassuring that these fluctuations are nevertheless both positive and negative and do not show up in all annuli.

\begin{figure}
\centering
\epsfig{file=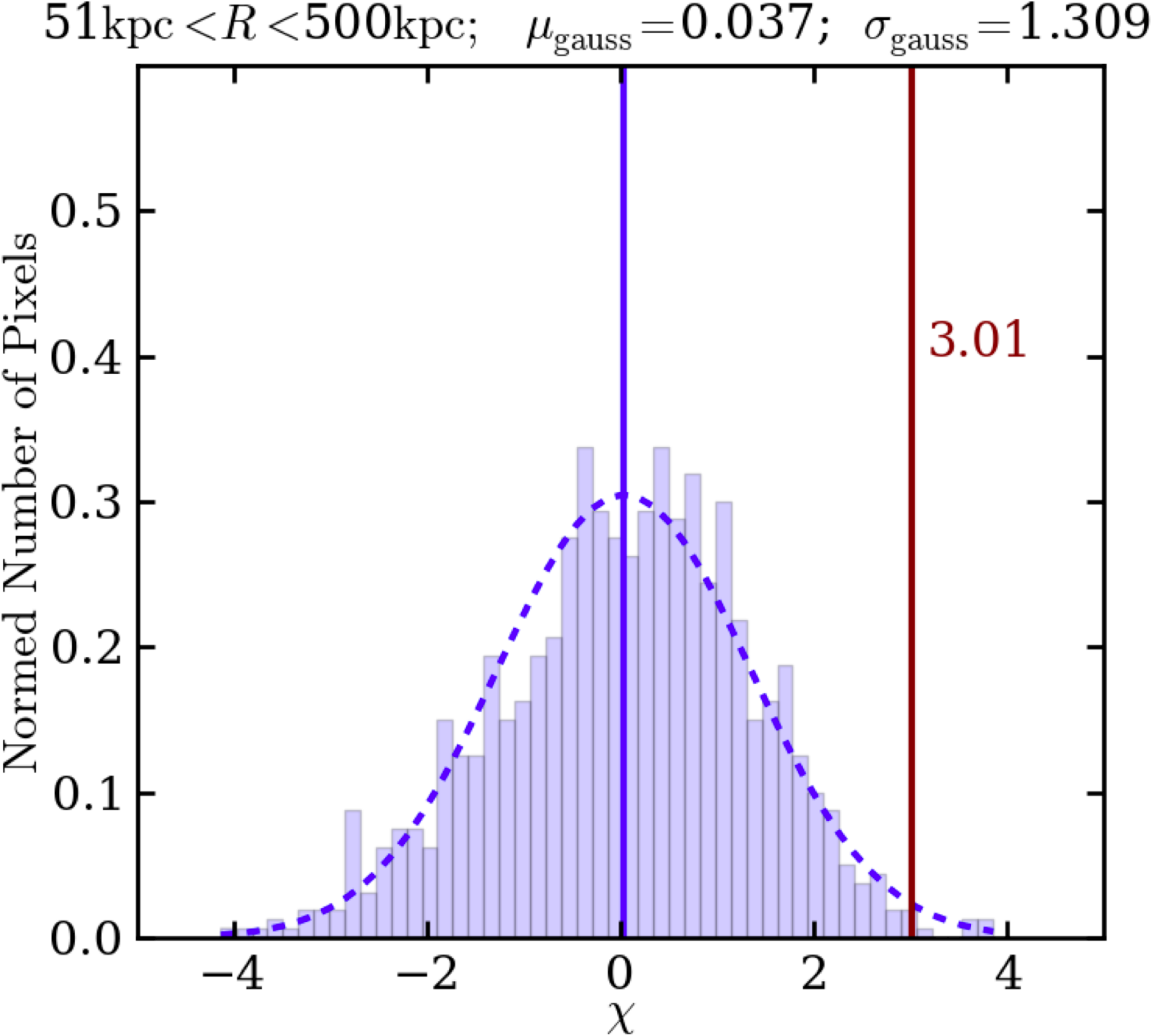, width=0.98\columnwidth} 
\caption[Statistical significance of our detection within 50 kpc and 500 kpc.]{$\chi$ histogram of the 1000 realizations 
of our same experiment using random samples of 15 stars each (see text for details) in a bin extending from $50$~kpc~to~$500$~kpc.
The red vertical line indicates the $\chi$ value for our measurement in the same bin.}
\label{MCMC_3}
\end{figure}

Further, although the statistical significance in a single radial bin is not high, we can combine our data to
increase the signal-to-noise ratio of our detection. 
We decide to consider the data in one large radial bin spanning from $50$~kpc~$<R<500$~kpc, and conduct a
similar Monte Carlo analysis.  Figure~\ref{MCMC_3} shows the $\chi$
histogram of the 1000 realizations in comparison to our measurement
(red vertical line) of the stacked PSF subtracted emission (i.e. QSO-star)
for this large radial bin. 
As expected from the previous analysis, the
distribution is clearly Gaussian and centered at zero\footnote{Note that in this case the mean value of the histogram of the Monte Carlo 
realizations is well centered on zero and consistent with the error on the mean.}, and we somewhat
underestimate the error.  Specifically, given that we find
$\sigma_{\rm gauss}=1.30$, our error is underestimated by 30\%.
Taking this into account, the deviation of our measurement ($\sim3\sigma$) from the distribution would 
correspond to a $2.32\sigma$ detection in this wide bin, or in other words, it falls in the 98.5th 
percentile of the Monte-Carlo realizations.

This detection corresponds to SB$_{\rm Ly\alpha}=(5.5\pm3.1)\times10^{-20}$~\unitcgssb within an annular region extending from $50$~kpc to $500$~kpc around the QSO. 
In the next section we will discuss the implications of this measurement in the context of
our current understanding of the gas distribution around QSOs, both from observations and 
from simulations.

\section{Discussion}
\label{sec:lastDisc}

As discussed in detail in the literature, given the resonant nature of the \lya line 
one needs to be particularly cautious when interpreting observations of this 
emission (e.g., \citealt{Gould1996, Cantalupo2005, Dijkstra2006, Verhamme2006}).
Further, as summarized in the introduction, multiple physical processes can potentially power extended 
\lya emission and disentangling them is challenging, especially given that several of them
could be at work.  These mechanisms are (i) resonant scattering of \lya line
photons (\citealt{Dijkstra2008,Steidel2011,Hayes2011}), (ii)
photoionization by stars or by a central AGN
(\citealt{Geach2009,Overzier2013}), (iii) cooling radiation from
cold-mode accretion (e.g.,
\citealt{Fardal2001,Yang2006,Faucher2010,Rosdahl12}), and (iv)
shock-heated gas by galactic superwinds (\citealt{Taniguchi2001,Dijkstra2012}).

In the following discussion we assume that the dominant mechanism
powering the faint Ly$\alpha$ emission we observed around quasars
is photoionization from the central AGN, resulting in a boosted \lya
fluorescence signal from the surrounding gas distribution.  Indeed,
scattering of \lya photons from the quasar should not be a major
contributor on such large scales ($>100$~kpc,
\citealt{Cantalupo2014}).  This is because the resonant scattering
process results in very efficient diffusion in velocity space, such
that the vast majority of resonantly scattered photons produced by the
quasar itself (or by the central galaxy) escape the system at very
small scales ($<10$~kpc), and hence do not propagate to larger
distances (e.g., \citealt{Dijkstra2006, Verhamme2006, Cantalupo2005,
  FAB2015}).  Further, in the case of \lya emission in a superwind we
would expect much higher values than observed for the \lya surface
brightness, i.e. at least SB$_{\rm Lya}\sim10^{-18}$~\unitcgssb
(\citealt{Taniguchi2001}), depending on shock velocity.  However, we
cannot rule out a cooling radiation scenario, and indeed, it is interesting to
note that the SB$_{\rm Ly\alpha}$ that we measure seems consistent
with the SB values shown for massive halos at large radii ($\gtrsim
100$~kpc) by \citet{Rosdahl12}, even though their simulations were run
to interpret the LAB phenomenon and do not include the presence of an
active AGN.  Despite this, they predict much higher emission (SB$_{\rm
  Lya}\sim10^{-18}$~\unitcgssb) at smaller scales ($R<100$~kpc) than
we observe.  Modeling the expected Ly$\alpha$ SB from cooling radiation clearly requires
further detailed investigation.

We thus interpret the extended emission that we measured as AGN
boosted Ly$\alpha$ fluorescence, and adopt two different approaches
for our analysis.  First, we consider the observed SB$_{\rm Ly\alpha}$
in the context of the simple model for cool halo gas described in
\citet{Hennawi2013} (already used in \citealt{FAB2015}), combined with
recent observational constraints on the cool gas distribution around
$z\sim2$ QSOs deduced from absorption line studies. Second, we compare
our emission measurement to the results of a cosmological hydrodynamical
simulation post-processed with ionizing and \lya radiative transfer, 
which was run to interpret the UM~287 nebula (see \citealt{Cantalupo2014}).

By studying absorption features in the spectra of projected quasar
pairs, the ``Quasars Probing Quasars'' series of papers (QPQ; \citealt{Hennawi2006}
and references thereafter) has shown that the optically thick
absorbers arising from gas clouds within the halo of quasars are
anisotropically distributed with respect to the quasar itself
(\citealt{Hennawi2007}). In other words, a quasar emits radiation
anisotropically or intermittently, so that the number of optically
thick absorption systems is higher in the background sightlines
(i.e. at a certain projected distance from the quasar) in comparison
to that observed along the line-of-sight to that quasar. This result,
together with the absence of fluorescent emission from these optically
thick absorbers (\citealt{Hennawi2013}) as well as photoionization modeling
of individual absorbers (\citealt{Prochaska2009,QPQ8}) all indicate a scenario in which
the optically thick gas observed in absorption toward background sightlines is shadowed from the
ionizing radiation of the quasar. \citet{Prochaska2013b} have shown that these optically thick
absorbers have a high covering factor $f_{\rm C}^{\rm thick}\simeq0.6$ on
$R\lesssim200$~kpc in the quasar's CGM \citep[see also][]{Hennawi2007,Prochaska2013,Hennawi2013}.

\citet{Hennawi2013} argued that, given this high covering factor and
the luminosity of the quasar, the optically thick clouds observed in absorption
would result in a very bright signal if directly illuminated by the ionizing radiation
of the quasar. We can estimate this SB for our sample
assuming the simple
model in \citet{Hennawi2013}, where a quasar is surrounded by a spherical halo
populated with spherical clouds of cool gas ($T \sim 10^4$\,K) at a
single uniform hydrogen volume density $n_{\rm H}$, which are uniformly
distributed throughout the halo, with an aggregate covering factor 
of $f_{\rm c}$.  By assuming this model, the \lya surface
brightness for an optically thick distribution of clouds is given by
(\citealt{Hennawi2013})
\bea
\label{SBthickValues}
{\rm SB}_{{\rm Ly}\alpha}^{\rm thick}&=& 6.7\times 10^{-17}\left(\frac{1+z}{3.253}\right)^{-4}\!\!
\left(\frac{f_{\rm C}^{\rm thick}}{0.5}\right)\!\!
\left(\frac{R}{160\,{\rm kpc}}\right)^{-2}\\ 
&\times&\left(\frac{L_{\nu_{\rm LL}}}{10^{30.9}\,{\rm erg\,s^{-1}\,Hz^{-1}}}\right)\cgssb\nonumber.
\eea
where we used the redshift of our observations ($z=2.253$), the virial radius of the dark matter halo hosting quasar as distance ($R=160$~kpc), 
and the specific luminosity at the Lyman limit (${{\rm log}\, L_{\nu_{\rm LL}}=30.9}$) corresponding to the average magnitude of our sample, i.e. $i$-mag~$=18.64$. 
This value of  $L_{\nu_{\rm LL}}$ is obtained as in \citet{FAB2015} by integrating the \citet{Lusso2015} composite 
quasar spectrum against the SDSS filter curve, and choosing the normalization which 
gives the correct magnitude for each quasar in our sample.
This signal is almost three orders of magnitude larger than the observed SB value of the \lya emission at 160~kpc 
from our stacking analysis, i.e. SB$_{\rm Ly\alpha}\sim10^{-19}$~\unitcgssb, and would have been
easily detected even in our individual exposures. Along the lines of the argument in \citet{Hennawi2013}, 
we conclude that the low SB signal that we observed cannot be arising from optically thick gas clouds illuminated by the quasar.

On the other hand, if we assume that gas with physical properties similar to the gas probed by absorption studies
is uniformly distributed
throughout the halo, the quasar would shine on the remaining $f_{\rm C}^{\rm thin} \simeq 1-f_{\rm C}^{\rm thick}$
component, keeping it
highly ionized, i.e. optically thin. Following \citet{Hennawi2013}, we can thus show that such optically thin clouds would result in 
a \lya SB level comparable to our observations

\bea
\label{SBthinValues}
{\rm SB}_{{\rm Ly}\alpha}^{\rm thin}&=& 8.8\times 10^{-20}\left(\frac{1+z}{3.253}\right)^{-4}\!\!
\left(\frac{n_{\rm H}}{0.01\,{\rm cm^{-3}}}\right)\\
&\times& \left(\frac{f_{\rm C}^{\rm thin}}{0.5}\right)\!\! 
\left(\frac{N_{\rm H}}{10^{20.5}\,{\rm cm^{-2}}}\right)\cgssb\nonumber, 
\eea
where we used the redshift of our observations ($z=2.253$), the median value for the column density $N_{\rm H}$ found in absorption studies (\citealt{QPQ8})
and an 
expected value for the volume density $n_{\rm H}$ from simulation on such scales (e.g. 160~kpc; \citealt{Rosdahl12}). 

Following the foregoing discussion, it is thus plausible that the low
\lya signal that we detect in our observations is due to optically
thin gas illuminated by the quasars.  Indeed, the only way to match
our observations via emission from optically thick clouds would be to
require a very low covering factor in eqn.~(\ref{SBthickValues}),
i.e. $f_{\rm C}^{\rm thick}\sim0.001$, but this would be clearly in conflict with
the results in \citet{Prochaska2013b}, arguing for the optically thin
scenario.  Note that optically thin emission is also in agreement with our previous
work on giant Ly$\alpha$ nebulae.  Indeed, although the quasars in our sample are on average
fainter  (average $i$-mag=18.64)
than the UM~287 quasar ($i$-mag=17.28,
\citealt{Cantalupo2014}), \citet{Jackpot} show that the $z\sim2$
quasar SDSSJ0841+3921 ($i$-mag=19.35), which is fainter than our average
quasar, is able to keep the gas highly ionized and optically thin out to scales of a few hundred kpc.
Specifically, SDSSJ0841+3921 is surrounded by a
large scale ($\sim300$~kpc) \lya nebulosity which is as
bright (SB$_{\rm Ly\alpha}=1.3\times10^{-17}$~\unitcgssb) \footnote{This is the average SB$_{\rm Ly\alpha}$ within the 2$\sigma$ 
S/N isophote, after removing the contribution from the unresolved emission of the 4 AGN.} as that surrounding
the UM~287 quasar (SB$_{\rm Ly\alpha}=1.2\times10^{-17}$~\unitcgssb) \footnote{This is the average SB$_{\rm Ly\alpha}$ within the 
same S/N isophote as for SDSSJ0841+3921, after removing the contribution from the unresolved emission of the 2 AGN.}.  Similar to UM~287, this \lya nebula is believed to result from
fluorescent emission powered by the ionizing radiation of the QSO.
Thus, the comparable SB$_{\rm Ly\alpha}$ in these two distinct cases, 
together with the
factor of $\sim20$ difference in the quasar ionizing luminosity\footnote{The ionizing luminosities are ${{\rm log}\, L_{\nu_{\rm LL}}}=31.73$ and 30.4, respectively for UM~287 and SDSSJ0841+3921
(\citealt{FAB2015, Hennawi2013}).}, strongly
suggests that brightness of the central source does not play an
important role, provided it is bright enough to ionize 
the surrounding gas distribution. This again favors an optically thin emission scenario,
since in this regime eqn.~(\ref{SBthinValues}) shows that
the \lya SB scales as 
SB$_{\rm Ly\alpha}\propto f_{\rm C}^{\rm thin}N_{\rm H}n_{\rm H}$, which is
independent of the luminosity of the quasar.

Having established that the optically thin regime is a good
hypothesis, we can now derive the physical properties of the emitting gas.
To break the
degeneracy between the covering factor $f_{\rm C}^{\rm thin}$, column density
$N_{\rm H}$, and volume density $n_{\rm H}$ intrinsic of our SB$_{\rm Ly\alpha}$ measurement in the optically thin regime, we
need independent constraints. 
Regarding $f_{\rm C}^{\rm thin}$, it has been shown that the smooth morphology of the emission in giant \lya nebulae implies 
a covering factor of $f_{\rm C}^{\rm thin}\gtrsim0.5$ (\citealt{FAB2015, FAB2015LAB, Jackpot}).
For simplicity, in what follows, we will always assume a covering factor of $f_{\rm C}^{\rm thin}=0.5$. 
Note that, as said before we are assuming $f_{\rm C}^{\rm thin} \simeq 1-f_{\rm C}^{\rm thick}$ , and 
thus this value is also well motivated by the distribution of optically thick absorbers, which show 
a high covering factor in the quasar's CGM (\citealt{Prochaska2013b}). 

Further, the total hydrogen column density $N_{\rm H}$ can be 
constrained via photoionization modeling of absorption systems
along the sightlines of background QSOs that pierce through the halo
of a foreground QSO (\citealt{QPQ8} and references therein).
In particular, \citet{QPQ8} compared photoionization models to a statistical sample of
absorbers in the CGM of typical $z\sim2$ QSOs, finding a median ${{\rm
    log}\, N_{\rm H}=20.5}$ within $200$~kpc from the quasars, although they
observed a substantial scatter in the distribution of $N_{\rm
  H}$ values, i.e. $\sim1.5$ dex (see Fig.~15 of \citealt{QPQ8}). 
A similar value  ${{\rm log}\, N_{\rm H}=20.4\pm0.4}$ (at impact parameter of $\sim180$~kpc)
was estimated from photoionization modeling of SDSSJ0841+3921, exploiting the so far
unique case where a giant \lya nebula could be modeled in both emission and absorption
(\citealt{Jackpot}). Given that these are the most recent and reliable estimates for $N_{\rm H}$ in the
literature which are also consistent with each other, we henceforth 
assume a value of ${{\rm
    log}\, N_{\rm H}=20.5}$ throughout our analysis.

Thus plugging in the values $z=2.253$, $f_{\rm C}=0.5$, and ${{\rm log}\, N_{\rm H}=20.5}$ into eqn.~(\ref{SBthinValues}), 
for the \lya SB in the optically thin scenario, we find the typical volume density $n_{\rm H}$ expected
on scales of ${\sim275}$~kpc \footnote{$R=275$~kpc is the average distance of the bin used in our analysis, i.e. $50$~kpc$<R<500$~kpc.},
given that SB${_{\rm Ly\alpha}=(5.5\pm1.8)\times10^{-20}}$~\unitcgssb from our measurement. 
We thus obtain an estimate of the volume density of the gas illuminated by the quasar and fluorescing in \lya:
\bea
\label{eqn:nHfound}
n_{\rm H} &=&0.6\times10^{-2}\left(\frac{{\rm SB}_{{\rm Ly}\alpha}^{\rm thin}}{5.5\times10^{-20}\cgssb}\right) \nonumber \\ 
&\times& \left(\frac{f_{\rm C}}{0.5}\right)^{-1}\left(\frac{N_{\rm H}}{10^{20.5}\,{\rm cm^{-2}}}\right)^{-1}\, {\rm cm}^{-3}.
\eea

Given this estimate for $n_{\rm H}$ at $R=275$~kpc, we can compute the ionization parameter $U \equiv \Phi_{LL}/c n_{\rm H} \propto L_{\nu_{\rm LL}}/n_{\rm H}$ 
at these distances around quasars. If we assume that $n_{\rm H}$ is not affected by illumination on these large scales, 
its value holds for both the shadowed and illuminated gas, and $U$ varies due to the different ionizing sources, either the
ultraviolet background (UVB; e.g., \citealt{HM2012}) if the gas is shadowed from the quasar emission, or the quasar if it is
illuminated. 
We find a value of ${\rm log}\, U_{\rm UVB}=-3.2$ and ${\rm log}\, U_{\rm QSO}=-0.2$ for shadowed gas and illuminated gas, respectively.
For randomly observed absorption systems, we thus expect to see a strongly bi-modal distribution of $U$ values.
Hence, it is interesting to compare this result with what was found by \citet{QPQ8} from photoionization modeling of individual
absorbers. Indeed photoionization modeling constrains the ionization parameter $U$, as long as ionic ratios can be measured (e.g. C$^{+}$/C$^{3+}$, Si$^{+}$/Si$^{3+}$). 
Note that the average ionizing luminosity of their sample is ${{\rm log}\, L_{\nu_{\rm LL}}=30.4}$, i.e. 0.5 dex lower than our sample, 
resulting in ${\rm log}\, U_{\rm QSO}=-0.7$.

In their data there is no sign for a bi-modal distribution in $U$ (see
Figure 7 in \citealt{QPQ8}), with low ionization parameter ${\rm
  log}\, U\lesssim-2$ at any distance from the quasar and no systems
at higher $U$, implying that these absorbers are likely not
illuminated by the quasar.
It is thus plausible that the Lau et
al. sample was biased against sources with very high $U$ values, which
would result in fewer detectable metal absorption lines needed to
constrain the ionization parameter through photoionization models.
Indeed, two systems out of 14 studied by \citet{QPQ8}, i.e. J0409-0411
and J0341+0000, do not have metal line detections, and thus do not
have constraints on the ionization state of the gas, i.e. no estimate
on $U$.
Nevertheless, the situation is more complicated with effects that can wash out the bi-modal 
distribution expected from the simple picture of half shadowed and half illuminated gas.
In particular, star-forming galaxies in the vicinity of the absorber could boost the $U$ value of shadowed sources from $U_{\rm UVB}$ to a higher $U_{\rm local}$. 
Further, the real distance between QSO and absorber is unknown, i.e. we have information only on the projected distance on the sky. 
If the real distance is 2-3 times larger than the projected distance, this would result in a reduction of 4-10 in $\Phi_{LL}$, and hence in $U$.  
In addition, if the highly ionized gas has similar velocity as the less ionized gas, its absorption features would then be lost in the stronger absorption 
of the less ionized gas, resulting in a bias against such high-U absorption systems.
Clearly, to confirm this scenario and test the relevance of these additional effects, further observations and analysis are needed.

However, given that we have an estimate for $U_{\rm QSO}$, it is then interesting to understand which kind of 
absorption features are expected from the gas that we detect here in emission
and compare them to the two systems in \citet{QPQ8}, which do not have metal line detections.
To estimate this, we use the photoionization code Cloudy (\citealt{Ferland2013}) to calculate the column densities expected for different ions for an optically thin
slab of gas at the distance of $R=275$~kpc from the quasar. As input parameters we used the aforementioned quantities that would give rise to the observed \lya emission, 
i.e. $N_{\rm H}=10^{20.5}$~cm$^{-2}$, ${\rm log}\, U_{\rm QSO}=-0.2$. We have then included the presence of the ultraviolet background (UVB; \citealt{HM2012}), and assumed the gas to have a metallicity of 
$Z=0.1$~Z$_{\odot}$, similar to values in \citet{QPQ8}.
This calculation shows that in the \lya emitting gas $N_{\rm HI}=10^{14.9}$~cm$^{-2}$, $N_{\rm CIV}=10^{12.8}$~cm$^{-2}$, $N_{\rm NV}=10^{12.9}$~cm$^{-2}$, 
$N_{\rm OVI}=10^{14.9}$~cm$^{-2}$, and both $N_{\rm SiII}$ and $N_{\rm FeII}$ have very low values ($\ll 10^{10}$~cm$^{-2}$), confirming that would be extremely difficult to
detect metal absorption from such systems.  
Indeed, the two highly ionized systems reported in \citet{QPQ8} have column densities comparable to these expectations. 
In particular, J0409-0411 shows $N_{\rm HI}=10^{14.2}$~cm$^{-2}$, $N_{\rm CIV}<10^{13.31}$~cm$^{-2}$, $N_{\rm SiII}<10^{13.48}$~cm$^{-2}$,
and has an impact parameter of 190 kpc from the quasar, while 
J0341+0000 similarly shows $N_{\rm HI}=10^{14.2}$~cm$^{-2}$, $N_{\rm CIV}<10^{13.28}$~cm$^{-2}$, $N_{\rm SiII}<10^{13.42}$~cm$^{-2}$, $N_{\rm FeII}<10^{13.62}$~cm$^{-2}$, 
and has an impact parameter of 235 kpc.
Thus, our analysis suggests that these two systems in \citet{QPQ8} are good candidates for background sightlines piercing gas which is illuminated by the quasar, 
and hence optically thin, and characterized by a high ionization parameter.

Lastly, we can compare our empirically derived $n_{\rm H}$ value with simulations. We are not aware of an average density profile for
the cool gas in massive (M$_{\rm DM}\sim10^{12.5}$~M$_{\odot}$) halos in the literature, although we
plan to determine it in future work. 
Simulations of massive halos (e.g., \citealt{Rosdahl12}) usually predict $n_{\rm H}$ values in the range ($\sim10^{-2}-10^{-3}$~cm$^{-3}$) on these scales ($\sim200$~kpc)
within filamentary structures. These values are in rough agreement with our estimate.
However, they may not optimally resolve this gas (\citealt{Cantalupo2014, Crighton2015}), and high densities are also probably present (\citealt{FAB2015, QPQ8}).

\citet{Cantalupo2014} showed that one can calibrate relations between
SB$_{\rm Ly\alpha}$ and the total hydrogen column density in
cosmological simulations post-processed with radiative transfer using
the RADAMESH Adaptive Mesh Refinement code (\citealt{Cantalupo2011}),
resulting in a picture which is consistent with the simple analytical
relations discussed above, i.e. eqn.~(\ref{SBthickValues}) and
eqn.~(\ref{SBthinValues}).  By comparing pixel by pixel the mock image
obtained for the SB$_{\rm Ly\alpha}$ and the $N_{\rm HII}$ column
density map of cool gas ($T<5\times10^4$~K) in the case where the gas
is mostly ionized by the quasar, i.e. what we above referred to as the
optically thin scenario, \citet{Cantalupo2014} obtained
the following relation\footnote{Note that in the highly ionized case
  $N_{\rm HII}$ is basically $N_{\rm H}$.} from their
simulation:
\begin{equation}
\label{Eqn:Seba}
N_{\rm HII} = 10^{21} \, \left(\frac{{\rm SB_{\rm Ly\alpha}}}{10^{-18}\, {\rm erg\, s^{-1}\, cm^{-2}\, arcsec^{-2}}}\right)^{1/2} C^{-1/2} \, {\rm cm^{-2}}.
\end{equation}
This relation depends on the clumping factor ${C=\langle n_{\rm H}^2 \rangle/\langle n_{\rm
    H}\rangle^2}$, introduced by \citet{Cantalupo2014} to account
for dense gas on small scales that are possibly unresolved by the simulation.
If we plug in our value of SB$_{\rm Ly\alpha}=5.5\times10^{-20}$~\unitcgssb, and assume $C=1$, we find ${{\rm log}\, N_{\rm H}=20.37}$, which is consistent with the simulation values (see Extended Data Fig.~3 of \citealt{Cantalupo2014}),
and in agreement, within the uncertainties, to the observational value we used in the foregoing discussion
${{\rm log}\, N_{\rm H}=20.5}$. 
We thus find agreement between simulations and our simple analytical model in the case of these low SB levels, 
as opposed to the case of the giant bright Ly$\alpha$ nebulae, where very high clumping factors are required by simulations to match the 
observed SB$_{\rm Ly\alpha}$, with values up to $C\sim1000$ (\citealt{Cantalupo2014}).

In summary, we argue that the hydrogen volume density around
typical QSOs on $> 100$~kpc scales
is approximately
constrained to be
$n_{\rm H} \simeq 0.6\times10^{-2}$~cm$^{-3}$, in broad agreement with the cool phase densities found in cosmological simulations
\footnote{Intriguingly this value  
is in
agreement with a radiation pressure confinement scenario (RPC; \citealt{Stern2014}), shown to be in place at least up to 10~kpc from an AGN.
In this scenario the radiation pressure is the 
main contributor to the gas pressure so that $P_{\rm rad}\sim P_{\rm gas}$, and the density profile is then shown to follow 
a simple relation. Following eqn.~(10) in \citealt{Stern2014}, 
using $R=275$~kpc, the average ionizing photon flux for our sample $Q_{\rm ion}\equiv 4\pi R^2 \Phi_{LL}\approx 10^{57}$~s$^{-1}$, 
$L_{i,45}=Q_{\rm ion} \langle h/\nu \rangle = 6\times10^{46}$~erg~$s^{-1}$, and $\tau_{\rm dust}=0.05$, we obtain
$n_{\rm H}=0.7\times10^{-2}$~cm$^{-3}$, remarkably close to what we found. 
However, to establish a RPC scenario on such large scales one would need a particular tuning of the time scales as the photons need to reach 300 kpc from the source. 
Nonetheless, it is worth noticing that any additional source of external pressure that would break the RPC scenario, would make the gas density higher (\citealt{Stern2014}).}.
However, bright giant \lya
nebulae, like UM~287 (\citealt{Cantalupo2014}) and SDSSJ0841+3921
(\citealt{Jackpot}), appear to require much higher densities $n_{\rm
  H}\gtrsim 1$~cm$^{-3}$ (\citealt{Cantalupo2014, FAB2015, Jackpot}),
to explain their higher SB \lya emission. 
Thus our analysis indicates a scenario in which
$\sim10$\% of quasars show bright 
giant \lya nebulae (SB$_{\rm Ly\alpha}\sim 10^{-17}$~\unitcgssb on scales $>50$~kpc) which require
high gas densities, while the remainder of the population is
characterized by fainter extended \lya emission 
and thus by much lower
gas densities.  At the moment, it is not clear what physical process
is responsible for these factor of $\sim 100$ difference in gas density. 
It has been argued that the
presence of a giant \lya nebula may be physically connected to the
location of over-densities of galaxies and AGN which provides an important
clue (\citealt{Jackpot}). 

\subsection{Comparison with Previous Deep Observations around Quasars}

Although many studies have targeted extended emission around
individual quasars (\citealt{HuCowie1987, Heckman1991spec,
  Heckman1991, Christensen2006, Hennawi2009, North2012, Hennawi2013,
  Roche2014, Herenz2015}), they do not achieve the SB sensitivity indicated by our
stacked profile of the quasar CGM. This is primarily due to the limited
sensitivity of the instruments employed in these studies or because
they had a different chief science goal, 
e.g. focused on studying the much brighter emission from the
EELR on smaller scales (\citealt{Christensen2006}).

On the other hand, both \citet{Rauch2008} and \citet{Cantalupo2012}
have targeted quasars down to interesting sensitivity levels that
should, according to our average profile, allow one to detect emission
on 100~kpc scales. In particular, \citet{Rauch2008} achieve a SB limit
of SB${_{\rm
    Ly\alpha}=8\times10^{-20}}$~erg\,s$^{-1}$\,cm$^{-2}$\,arcsec$^{-2}$
($1\sigma$ in 1 arcsec$^2$) in their 92~h long-slit observations of
the quasar DMS 2139-0405 ($z=3.2$; \citealt{Hall1996}), while
\citet{Cantalupo2012} reach SB${_{\rm
    Ly\alpha}\sim4\times10^{-19}}$~erg\,s$^{-1}$\,cm$^{-2}$\,arcsec$^{-2}$
($1\sigma$ in 1 arcsec$^2$) with 20~h of narrow-band imaging targeting
the $z=2.4$ quasar HE0109-3518.  Although these studies reached
interesting depths, they do not appear to show evidence for any
extended \lya emission on large scales.  However, as we have shown,
probing these extremely low surface brightness levels on 100~kpc
scales requires a very careful analysis with proper accounting of
systematics (e.g., flat-fielding, sky and continuum subtraction,
contamination from nearby sources) and a specifically tailored
analysis. It would be interesting to search for the extremely
faint CGM Ly$\alpha$ emission that we detected in our stack 
in these very deep observations of individual quasars, which
we plan to pursue in future work.

\subsection{Comparison with LBGs and LAEs \lya Profiles}
\label{sec:comp}

\begin{figure*}
\centering
\epsfig{file=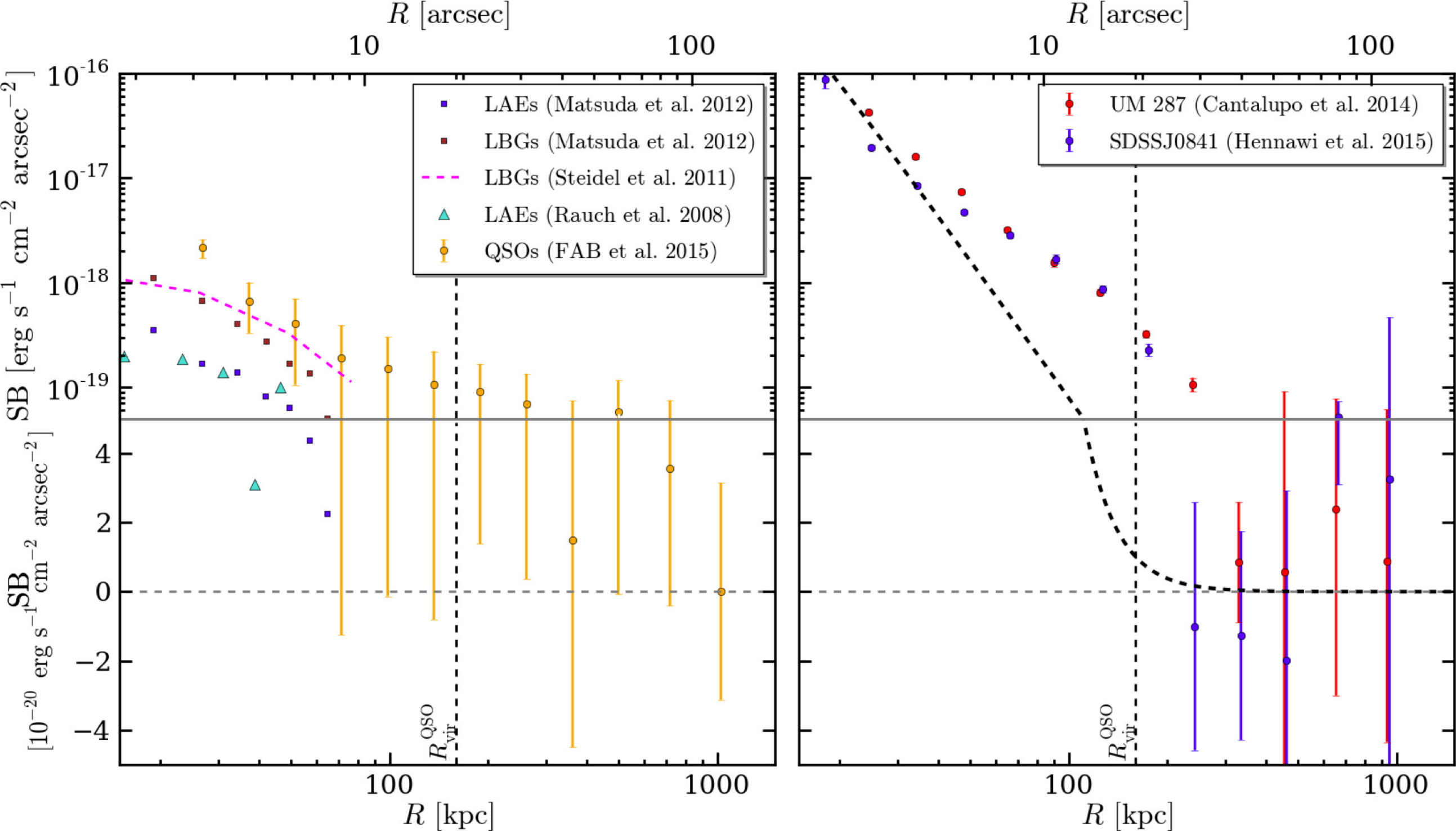, width=1.17\columnwidth, angle=270, clip} 
\caption{Comparison of the extended \lya surface brightness profiles for QSOs, LAEs, and LBGs. {\bf Left panel:} We compare the profile already shown in Figure~\ref{comp_profiles_norm} with profiles reported in the literature: 
average profile of 92 continuum-selected LBGs at $\langle z \rangle=2.65$ (dashed magenta line; \citealt{Steidel2011}), average profile of 24 proto-cluster LBGs at $z\sim3$ (brown squares; \citealt{Matsuda2012}), 
average profile of 130 LAEs at $z\sim3$ (blue squares; \citealt{Matsuda2012}), median profile of 27 faint line emitters at $2.66 < z < 3.75$ (cyan triangles; \citealt{Rauch2008}). 
{\bf Right panel:} We show the \lya surface brightness profile for 
the giant nebulae around UM~287 (red; \citealt{Cantalupo2014})and around SDSSJ0841+3921 (blue; \citealt{Jackpot}). Note that these two profiles are not PSF subtracted as in our stacking analysis. For this reason,
we show a Moffat profile (black dashed curve) representing the seeing of the observations.
In both panels, the vertical dashed line indicates the expected virial radius for a QSO at $z\sim2$, i.e. $R_{\rm vir}^{\rm QSO}\sim160$~kpc (\citealt{Prochaska2014}).
Note that all QSO profiles (the one presented in this work, around UM~287 and SDSSJ0841+3921) have a signal of SB$_{{\rm Ly}\alpha}\sim10^{-19}$ \unitcgssb at 200~kpc.}
\label{Comp_prof}
\end{figure*}

With the aim of comparing our newly determined \lya profile with
previous observations at comparable redshift, and to gain a better
overview on extended \lya halos, we compile extended \lya profiles for
different sources from the literature.
The left panel of Figure~\ref{Comp_prof}
shows our data together with the average profile of 92
continuum-selected LBGs at $\langle z \rangle=2.65$ (dashed magenta
line; \citealt{Steidel2011}), the average profile of 24 proto-cluster
LBGs at $z\sim3$ (brown squares; \citealt{Matsuda2012}), the average
profile of 130 LAEs at $z\sim3$, selected to be in dense regions
$2.5<\delta_{\rm LAE}<5.5$ (blue squares; \citealt{Matsuda2012}), and
the median profile of 27 faint line emitters at $2.66 < z < 3.75$
(cyan triangles; \citealt{Rauch2008}).

The origin of \lya halos around these star-forming galaxies is still a
matter of debate, and, as for the case of giant \lya nebulae around
quasar, several mechanisms have been discussed in the literature.
First, it is important to point out that \lya cooling radiation is not
able to account for the high \lya luminosity of these halos,
i.e. $L_{\rm Ly\alpha}\gtrsim 10^{42}$~erg~s$^{-1}$, giving rise to
halos which are an order of magnitude fainter than observed (e.g.,
\citealt{Haiman2000,Dijkstra&Loeb2009, Dijkstra2012}, but see also
\citealt{Rosdahl12}).  However, an additional boost to the \lya
emission could result from the ionizing photons emitted by the galaxy
itself or by nearby galaxies (e.g., \citealt{Dijkstra&Loeb2009, MasRibas2016}).
The photoionized gas could then emit enough \lya photons to be consistent
with the observations. This would depend, however, on the ionizing photon
escape fraction.
Further, resonant scattering is {\it not} able to reproduce the extent of the \lya
halo, giving rise to more compact emission than observed (e.g.,
\citealt{Dijkstra2012, Lake2015}), unless bipolar outflows in which the
clumps decelerate (in the case of LBGs; \citealt{Dijkstra2012}), or
additional sources of \lya emission (in the case of LAEs;
\citealt{Lake2015}), such as low star-formation
spatially distributed inside the host dark matter halo or cooling
radiation, are taken into account.
In summary, current studies suggest that in the
case of LAEs and LBGs, as it is the case for QSO, a strong
contribution from photoionization is needed in order to explain the
extent and surface brightness of the \lya halo observed, 
with up to 50-60\% of the observed \lya signal due to this powering mechanism (\citealt{MasRibas2016}).

In the right panel of Figure~\ref{Comp_prof},
we show the \lya profile of the UM~287 nebula
(red; \citealt{Cantalupo2014}) and of the \lya nebula around
SDSSJ0841+3921 (blue; \citealt{Jackpot}) that we have computed from
the continuum-subtracted NB image by averaging in circular apertures
around the QSO after masking all the sources from the $V$-band
(i.e. available broad band) and compact sources from the NB, as done
for our individual objects in the GMOS observations (see
\S\ref{imPrep}). Note that the data reduction of UM~287 and
SDSSJ0841+3921 has been performed differently. First of all, the \lya
image has been obtained through a continuum-subtraction (see
\citealt{Cantalupo2014, Jackpot}), and {\it no} empirical
PSF-subtraction has been performed on these data.  Further, no
detailed variance images have been computed. For these reasons, the
errors shown in this plot for UM~287 and SDSSJ0841+3921 do not include
the errors on the continuum and PSF subtraction, but are instead
estimated through a bootstrap analysis ($1\sigma$). Note that these
profiles, as highlighted in our analysis, could be affected by
systematics at large radii (i.e. $\gtrsim200$~kpc), where
imperfections in the data reduction and in the data itself may be
dominant (e.g., flat-fielding, sky-subtraction, continuum
subtraction).
In both panels we indicate with a vertical line the expected virial radius  
for a QSO at this redshift ($R_{\rm vir}^{\rm QSO}\sim160$~kpc, \citealt{Prochaska2014}). As stressed already in our previous work,
the extent of the \lya emission goes beyond the expected size of the dark matter halo hosting quasars.

Figure~\ref{Comp_prof} shows that QSOs and LBGs have a higher emission profile which extends further out 
than LAEs.
This could reflect the physics behind the \lya emission, 
with objects able to ionize a larger amount of gas out to larger distances
and characterized by a denser 
environment surrounded by higher and more extended \lya signal.
With the amount and distribution (density) of gas probably determining the strength of the \lya emission (i.e. optically thin scenario), and with the 
environment probably determining the extent (\citealt{Matsuda2012}).
In other words, this plot seems to suggest that more massive systems, i.e. the quasar population, show
higher \lya surface brightness profiles because the stronger ionizing radiation
from the central source (or from nearby galaxies)\footnote{Note that seems unlikely that unresolved/undetected galaxies could have a large contribution to the 
\lya emission that we
detect. Indeed, if we use the flux limit of $f_{\rm Ly\alpha}=3.0\times10^{-17}$~erg~s$^{-1}$~cm$^{-2}$ (above which we are surely complete), we get 
a crude estimate of the star-formation rate SFR$\approx1$~M$_{\odot}$~yr$^{-1}$ for undetected galaxies in our NB data, 
using the formula in \citet{Kennicutt1998} and case B. Satellites with SFR of this order 
seem to have a small impact on the brightness and morphology of extended \lya emission on scales of $100$~kpc (\citealt{Cen2013}).}
is able to ionize the gas at much larger distances than in less massive systems.
Indeed, `average' LBGs and LAEs rely only on the
ionizing radiation of stars (i.e. no bright QSO or AGN) and populate lower
mass halos, i.e. $M_{\rm DM}\sim 10^{12}$~M$_{\odot}$ and $M_{\rm DM}\sim 10^{11}-10^{11.5}$~M$_{\odot}$ for LBGs and LAEs, respectively. 
Thus, probably for these reasons, their \lya profile is
lower and less extended.

On the other hand, systems like UM~287 and SDSSJ0841+3921 probably represent  
rare specific cases (i.e. $\sim10$~\% of quasars) where the gas densities are 
simply much higher. 
However, it is interesting to note that at $\sim200$~kpc, both our stacked
profile and the profiles for UM~287 and SDSSJ0841+3921 show a \lya signal of SB${_{\rm
    Ly\alpha}\sim 10^{-19}}$~\unitcgssb. It is thus plausible that our
stacking analysis is consistent with (or detects) the same gas in
the CGM or filamentary structures on these large scales around the quasars
as for UM~287 and SDSSJ0841+3921 (\citealt{Cantalupo2014, FAB2015, Jackpot}).  However, deep observations of a
large sample of QSOs are needed to firmly characterize their extended
\lya profile with the same precision as for the LBGs and LAEs,
and
conclusions could be drawn only after disentangling the several
mechanisms that can produce low SB \lya emission.

\section{Summary and Conclusions}
\label{sec:Conclusion}

Using NB imaging data collected as part of the FLASHLIGHT-GMOS survey (see also \citealt{FAB2013}), 
we have performed a stacking analysis to characterize the extended \lya emission around typical bright $z\sim2$ QSOs.
We find that:

\begin{itemize}

\item The 15 QSOs in our sample do not show giant \lya nebulae similar
  to UM~287 (\citealt{Cantalupo2014}) or SDSSJ0841+3921
  (\citealt{Jackpot}), i.e. SB$_{{\rm Ly}\alpha} =
  10^{-17}$ \unitcgssb
  emission on scales $> 50$~kpc, even though
  we would have been able to detect such emission.  The
  PSF subtraction reveals \lya emission above SB$_{{\rm Ly}\alpha} =
  10^{-17}$ \unitcgssb extending on scales of radius $\sim 3-4$ arcsec
  ($\sim 25-32$~kpc) around 7 out of 15 quasars, i.e. 47\% of the sample.

\item The average radial profile of the 15 QSOs in our sample shows a deviation from the Moffat PSF of our NB images,
starting at $\sim70$~kpc at around SB${_{\rm Ly\alpha}\sim10^{-19}}$ erg\,s$^{-1}$\,cm$^{-2}$\,arcsec$^{-2}$. This can be translated to a low significance first 
radial profile of the \lya emission of the quasar CGM (see Figure~\ref{comp_profiles_norm}).
Using a Monte Carlo analysis (see \S\ref{sec:Monte Carlo}), we ascertain that we have a $\sim2.32\sigma$ detection 
within an annular bin spanning $50$~kpc~$<R<500$~kpc from the QSOs. The \lya emission in this bin, centered at 
$R=275$~kpc, is estimated to be SB${_{\rm Ly\alpha}=(5.5\pm3.1)\times10^{-20}}$~erg\,s$^{-1}$\,cm$^{-2}$\,arcsec$^{-2}$.

\item Assuming a scenario in which the illuminated gas is highly ionized by the quasar radiation,
  the detected SB$_{\rm
  Ly\alpha}$ on scales of hundreds of kpc implies gas volume densities
    of the order $n_{\rm H}=0.6\times10^{-2}$~cm$^{-3}$.  This value
    is much lower than what has been proposed in the case of extended
    bright \lya nebulae around UM~287 (\citealt{FAB2015}) and
    SDSSJ0841+3921 (\citealt{Jackpot}), i.e. $n_{\rm H}\gtrsim
    1$~cm$^{-3}$.  However, our stacking analysis results in a \lya
    profile consistent with the \lya profile of UM~287 and
    SDSSJ0841+3921 on larger scales of $\sim200$~kpc (SB${_{\rm
        Ly\alpha}\sim 10^{-19}}$~\unitcgssb).
    Thus on these larger scales, it plausible that our stacking analysis detect a signal from
    the CGM or IGM as has been observed around UM~287 and SDSSJ0841+3921.

\end{itemize}

Future surveys targeting the \lya line around QSOs with new IFU instruments, i.e. MUSE (\citealt{Bacon2010}) and KCWI (\citealt{Morrissey2012}),
will be able to investigate the CGM of quasars without the tight restrictions on redshift range and redshift accuracy inherent in narrow-band imaging.
This will allow one to observe a much broader range of quasars, enabling studies of quasars with specific attributes such as hyper-luminous quasars,  obscured quasars,
radio-loud quasars, or quasars in richer environments.  Such studies are needed to test our current interpretation and to determine if specific types of quasar preferentially exhibit
bright emission from their CGM.

Further, the lack of bright giant \lya nebulae in our data confirm the necessity of large statistical samples of QSOs 
to uncover the brightest \lya nebulae on the sky.
The study of these bright nebulae, such as the UM~287 nebula (\citealt{Cantalupo2014}) or SDSSJ0841+3921 (\citealt{Jackpot}), 
enable further  characterization of the CGM gas in individual systems via the detection and modeling
of additional emission
lines, e.g. \civ, \heii (\citealt{FAB2015LAB, FAB2015}).
The resulting physical properties can then be used as additional
constraints on simulations of galaxy-formation.
Indeed, we call attention to the lack of high-resolution cosmological zoom simulations of very massive systems ($M_{\rm halo}\sim10^{12.5}$~M$_{\odot}$) to which one could 
compare our results (but see \citealt{Fumagalli2014} and \citealt{Faucher-Giguere2016} for recent progress on this front).
Given the complexity of the radiative transfer of the resonant \lya line, 
simulations with both ionizing and \lya radiative transfer of massive halos are needed to fully capture the physics in play
in giant \lya nebulae, and to compare to the wealth of observational data that are now becoming available.

\acknowledgments 

We thank the members of the ENIGMA group\footnote{http://www.mpia-hd.mpg.de/ENIGMA/} at the
Max Planck Institute for Astronomy (MPIA) for helpful discussions.
In particular, FAB warmly thanks Jonathan Stern, Emanuele P. Farina, and Daniele Sorini for 
useful comments on an early draft of this work.
JXP acknowledge  support  from  the  National  Science  Foundation  (NSF)  grant  AST-1010004,
AST-1109452, AST-1109447 and AST-1412981.

\bibliographystyle{apj}
\bibliography{biblio}

\clearpage

\appendix

\section{Estimate of the Flux Losses due to the Error on the Redshifts}

As already stated in Section \S\ref{sec:data}, the typical uncertainty on the redshift of our sample is 
$\sigma_{z}\sim0.003$ (or equivalently $\sim270$~km~s$^{-1}$), which is much smaller than the width of 
the narrow-band filter used, i.e. $\Delta z=0.027$ (or equivalently $\Delta v=2479$~km~s$^{-1}$). 
Further, to be sure that the \lya emission of our targets fall within the narrow-band filter, we selected only
QSOs whose redshift gives a maximum shift of $\pm5$~\AA\ from the filter's center (or equivalently $\delta v = 370$~km~s$^{-1}$).

Nevertheless here, we test the robustness of our results against errors on the systemic redshift of the quasars in our sample. 
In particular, we estimate the flux losses due to this uncertainty. We proceed as follows. 
First, starting from the known systemic redshift of our sample, we build 1000 realizations of our 15 quasars 
by assuming that the error on the redshift is described by a gaussian distribution with $\sigma_{v}=300$~km~s$^{-1}$, 
$750$~km~s$^{-1}$, $1500$~km~s$^{-1}$, or $3000$~km~s$^{-1}$. Second, we assume that the signal we are looking for, i.e. 
the extended \lya emission on scale of the CGM, is characterize by a ${\rm FWHM}=600$~km~s$^{-1}$ centered at the systemic redshift of the quasar, 
as in the case of the giant nebula around UM~287 (\citealt{FAB2015}). Knowing the shape of our narrow-band filter, we are then able
to compute the amount of flux lost for each sample of the 1000 realizations by taking the average of the loss per quasar within each sample.

\begin{figure}
\centering
\epsfig{file=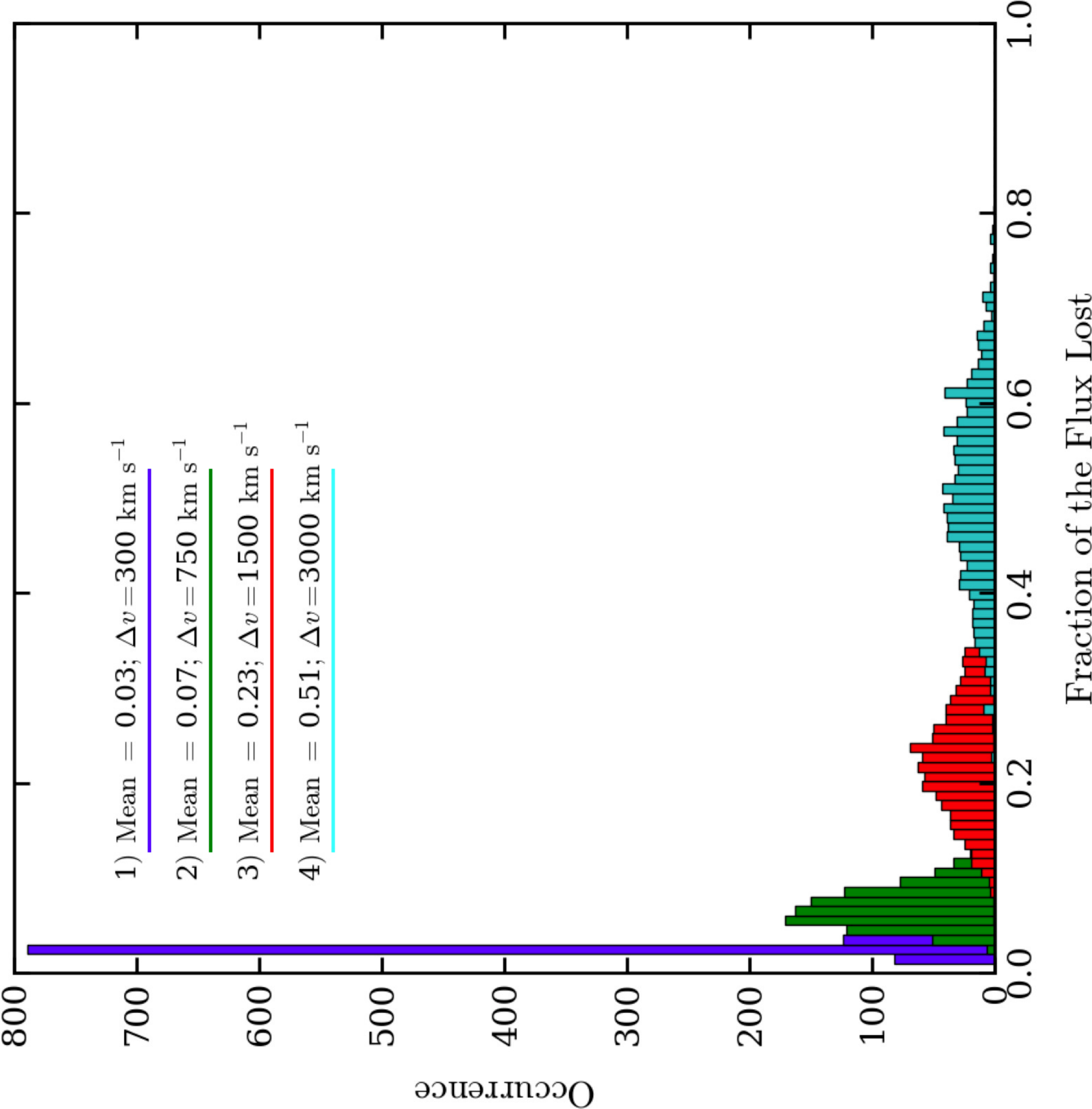, width=0.55\columnwidth, angle=270, clip} 
\caption{Estimate of the flux losses due to the error on the redshifts of our sample. Each histogram shows the distribution of the average flux 
loss for 1000 samples of 15 QSOs whose error on the redshift is assumed to be gaussian distributed with a $\sigma_{v}=300$~km~s$^{-1}$ (blue), 
750~km~s$^{-1}$ (green), 1500~km~s$^{-1}$ (red), or 3000~km~s$^{-1}$ (cyan). Given that the errors on the redshift of our sample are of the order of 300~km~s$^{-1}$, the flux
losses do not affect our measurement.}
\label{Flux_loss}
\end{figure}

Figure~\ref{Flux_loss} shows the distribution of these average flux losses. This test shows that our sample should only be affected by 
a flux loss of the order of 3\% (see blue histogram), while even doubling the error on the redshift would result in a 7\% loss (green histogram).
Only error distributions with width comparable to the width of our narrow-band filter can result in significant flux losses (see red and cyan histograms), and
thus alter remarkably our analysis.
Given the small uncertainty on the redshifts of our quasars, and our 
conservative selection criteria, we conclude that our sample do not suffer from flux losses.

\end{document}